# Long-term Simultaneous Monitoring Observations of SiO and H$_2$O Masers toward the Mira Variable WX Serpentis

Jang-Ho Lim[1,2], Jaeheon Kim[1], Se-Hyung Cho[3], Hyosun Kim[1], Dong-Hwan Yoon[1], Seong-Min Son[2], and Kyung-Won Suh[2]
[1] Radio Astronomy Division, Korea Astronomy and Space Science Institute, Yuseong-gu, Daejeon 34055, Republic of Korea; jhkim@kasi.re.kr
[2] Department of Astronomy and Space Science, Chungbuk National University, Cheongju, 28644, Republic of Korea
[3] Seoul National University, Gwanak-gu, Seoul 08826, Republic of Korea

## Abstract

We present the results from long-term simultaneous monitoring observations of SiO and H$_2$O masers toward the Mira variable star WX Serpentis. This study has been conducted with 21 m single-dish radio telescopes of the Korean VLBI Network from 2009 June to 2021 June. Five maser lines were considered: SiO $v=1, 2, J=1-0$; SiO $v=1, J=2-1, 3-2$; and H$_2$O $6_{1,6}-5_{2,3}$ transitions, with the SiO maser lines distributed near the stellar velocity and the H$_2$O maser exhibiting an asymmetric line profile with five to six peaked components. Intense H$_2$O maser emissions suddenly appeared in 2019 September, indicating flaring. The intensity variations of SiO and H$_2$O masers are strongly correlated with the optical light curve (OLC) of the central star, with individual phase lags; the phase lag of the H$_2$O maser relative to the OLC is larger than that of the SiO masers. The consequent phase difference between the SiO masers and the H$_2$O maser likely indicates that their formation regions and main driving mechanisms are different from each other. The SiO masers in WX Ser exhibit a dominant single-peak velocity distribution, similar to other Mira variable stars. However, the H$_2$O maser displays distinct morphological features, showing a radial acceleration and preferential intensity dominance at blueshifted velocities. This suggests that the H$_2$O maser clouds of WX Ser are moving outward, thereby developing an asymmetric outflow owing to nonuniform material ejection from the stellar atmosphere. The findings confirm that an initial asymmetric outflow structure emerged during the thermally pulsing asymptotic giant branch phase, specifically in the Mira variable star stage.

*Unified Astronomy Thesaurus concepts:* Silicon monoxide masers (1458); Water masers (1790); Asymptotic giant branch stars (2100); Circumstellar envelopes (237); Time series analysis (1916)

*Materials only available in the online version of record:* figure set, machine-readable tables

## 1. Introduction

During the asymptotic giant branch (AGB) phase in the Hertzsprung–Russell diagram, stars with low and intermediate initial masses ($\sim 1-8\ M_\odot$) deplete their hydrogen fuel through nuclear reactions. In this process, stellar materials are expelled into the interstellar medium ($\sim 10^{-8}$ to $10^{-4}\ M_\odot\ \mathrm{yr}^{-1}$) through a series of repeated events involving the contraction and expansion of the core and H-burning shell. These events play a crucial role in determining the fate of these stars (M. Jura & S. G. Kleinmann 1989). In general, oxygen-rich AGB stars exhibit periodic variations in temperature and radius over a span of $\sim$100–1000 days. These variations result in the ejection of stellar material, which eventually forms circumstellar envelopes (CSEs) composed of gas and dust, including oxygen atoms (S. Höfner & H. Olofsson 2018). In addition, these conditions promote the formation of molecules, such as SiO, H$_2$O, and OH, and increase their density levels to be sufficient for maser generations. Thus, the observation of maser lines can provide valuable insights into the kinematics and structure of the stellar atmosphere and CSE.

In stars with abundant oxygen, a stratified structure consisting of SiO, H$_2$O, and OH masers is often observed, extending from the stellar atmosphere to the edge of the CSE. The SiO maser region is located at a distance of $\sim$2–4 stellar radii ($R_*$) from the photosphere, positioned between the photosphere and dust layer (P. J. Diamond et al. 1994). Here the "photosphere" is defined as a continuum-forming layer in the optical and near-IR (M. D. Gray et al. 2009). Under the influence of periodic stellar pulsations and gravity, SiO masers exhibit complex kinematic movements (P. J. Diamond & A. J. Kemball 2003). H$_2$O masers exhibit features that indicate radial acceleration in the region between $5R_*$ and $20R_*$ from the photosphere (J. A. Yates & R. J. Cohen 1994; A. M. S. Richards et al. 2011). OH masers are typically observed in the outermost part of the CSE (beyond $100R_*$), where the wind velocity acceleration process has stopped (I. Bains et al. 2003).

The Mira variable star WX Serpentis (=IRAS 15255+1944, IRC+20281, CIT 7) exhibits a period of approximately 425 days, based on optical data provided by the American Association of Variable Star Observation (AAVSO). The spectral classification of this object is reported to be between M8.5IIIe and M10III (W. Buscombe & B. E. Foster 1999). T. J. Jones et al. (1988) reported that this object shows the characteristics of a typical Mira variable star from the infrared properties. The distance to WX Ser has been determined to be approximately 756 pc, based on annual parallax measurements obtained from the Gaia satellite (EDR3; Gaia Collaboration 2022). The rate of mass loss has been estimated to be $\sim 5.3 \times 10^{-7} \, M_\odot \, yr^{-1}$ through CO molecular line observations (G. R. Knapp 1986). Initial observations of maser emissions originating from SiO, $H_2O$, and OH molecules in WX Ser were made by W. J. Wilson & A. H. Barrett (1972), L. E. Snyder & D. Buhl (1975), and D. F. Dickinson (1976), respectively. Subsequent observations of SiO and $H_2O$ maser lines in WX Ser have been conducted as part of various survey projects (M. Balister et al. 1977; S.-H. Cho et al. 1998; J. Kim et al. 2010; J. R. Rizzo et al. 2021). Although monitoring observations of OH and $H_2O$ maser lines have been performed, the durations were limited to 2 and 3 yr, respectively (J. Herman & H. J. Habing 1985; M. Shintani et al. 2008). In the case of SiO masers, no systematic monitoring has been documented. Consequently, the maser properties associated with the Mira variable star WX Ser have not yet been comprehensively explored, resulting in limited knowledge regarding the correlation of SiO and $H_2O$ masers with the optical light curve (OLC) of WX Ser and the corresponding characteristics. This research gap prompted us to conduct simultaneous, long-term monitoring observations of SiO and $H_2O$ masers, using the single-dish telescope of the Korean VLBI Network (KVN). The KVN is equipped with four-band receivers ($K$, $Q$, $W$, and $D$), enabling the simultaneous observations of SiO and $H_2O$ masers, as described by S.-T. Han et al. (2013).

**Table 1**
Observed Molecular Transitions and Rest Frequencies

| Molecule | Transition | Frequency (GHz) | Abbreviation | Detected | Epochs |
| --- | --- | --- | --- | --- | --- |
| $^{28}$SiO | $v = 0, J = 1-0$ | 43.423 | ... | × | 1 |
|  | $v = 0, J = 2-1$ | 86.846 | ... | × | 1 |
|  | $v = 1, J = 1-0$ | 43.122 | SiO43 | ○ | 50 |
|  | $v = 1, J = 2-1$ | 86.243 | SiO86 | ○ | 41 |
|  | $v = 1, J = 3-2$ | 129.363 | SiO129 | ○ | 36 |
|  | $v = 2, J = 1-0$ | 42.820 | SiO42 | ○ | 50 |
|  | $v = 2, J = 2-1$ | 85.640 | ... | × | 1 |
|  | $v = 2, J = 3-2$ | 128.458 | ... | × | 1 |
| $^{29}$SiO | $v = 0, J = 1-0$ | 42.879 | ... | ○ | 7 |
| $H_2O$ | $6_{1,6}$–$5_{2,3}$ | 22.235 | $H_2O$ | ○ | 50 |

This paper presents the results of long-term simultaneous monitoring observations. The observations are outlined in Section 2, and the results are presented in Section 3. The results are comprehensively discussed in Section 4. Section 5 presents the concluding remarks.

## 2. Observations

Over a 12 yr period, i.e., from 2009 June to 2021 June, we carried out simultaneous monitoring observations of SiO and $H_2O$ masers toward WX Ser using three KVN 21 m single-dish telescopes (KVN Yonsei, KVN Ulsan, and KVN Tamna). The observation covered a broad frequency range, including 10 transitions of the SiO and $H_2O$ isotopes. The center frequencies of these transition lines are documented in Table 1. Whether the line has been detected with the 3$\sigma$ level is marked in the fifth column, and the number of observations is given in the sixth column.

During 2009–2012, only $K$- and $Q$-band observations were available, when the initial phase of our monitoring was conducted. Later, in 2012 May, the development of the $W$- and $D$-band receivers was completed and the KVN four-band receiver system commenced operations (S.-T. Han et al. 2013). Consequently, we observed the $H_2O$ maser at 22.235 GHz (hereafter $H_2O$); $^{28}$SiO $v = 1, J = 1-0$ at 43.122 GHz (hereafter SiO43); $^{28}$SiO $v = 2, J = 1-0$ at 42.820 GHz (hereafter SiO42) throughout the entire observation period (2009–2021); and $^{29}$SiO $v = 0, J = 1-0$ at 42.879 GHz in seven epochs during 2009–2012. Since 2012, other lines were included in our monitoring program, i.e., $^{28}$SiO $v = 1, J = 2-1$ at 86.243 GHz (hereafter SiO86) and $^{28}$SiO $v = 1, J = 3-2$ at 129.363 GHz (hereafter SiO129), as well as two thermal lines; $^{28}$SiO $v = 0, J = 1-0$ and $2-1$ lines; and $^{28}$SiO $v = 2, J = 2-1$ and $3-2$ maser lines, which are known to be very weak and poorly detected (however, these four lines were only observed once to investigate their detectability).

To summarize, we collected a total of 50 epochs of data for SiO43, SiO42, and $H_2O$ maser lines throughout the observation period (2009 June–2021 June). Additionally, we obtained 41 epochs of data for the SiO86 maser and 36 epochs of data for the SiO129 maser between 2012 May and 2021 June. The monitoring observations were conducted at regular intervals of 2–5 months, except for two instances in which the observations were performed over a duration of 1 yr each, coinciding with the implementation of other projects and an upgrade of the telescope system.

This study constitutes an ongoing research endeavor building on the previous investigations conducted by J. Kim et al. (2016, 2019) and D.-H. Yoon et al. (2023). It is a part of comprehensive, long-term monitoring programs focused on SiO and $H_2O$ masers targeting AGB stars, using the KVN radio telescopes. To maintain consistency with previously conducted studies, we exclusively used left-circular polarization for our observations. From 2009 to 2012, a 4096-channel digital filter bank (DFB) was used. The DFB had a combined bandwidth of $4 \times 64$ MHz and was adopted for the $K$ and $Q$ bands (SiO43; SiO42; $^{29}$SiO $v = 0, J = 1-0$; and $H_2O$ maser lines). The velocity resolutions were approximately 0.21 km s$^{-1}$ ($K$) and 0.11 km s$^{-1}$ ($Q$). Following the implementation of the upgraded simultaneous four-band system in 2012 June, the total bandwidths for the $K$ and $Q$ bands were $4 \times 32$ MHz, and those for the $W$ and $D$ bands (SiO86 and SiO129 maser lines) were $2 \times 64$ MHz. The velocity resolutions were approximately 0.11 km s$^{-1}$ ($K$), 0.05 km s$^{-1}$ ($Q$), 0.05 km s$^{-1}$ ($W$), and 0.036 km s$^{-1}$ ($D$). The Hanning smoothing technique was applied to all spectra to ensure a uniform velocity resolution of $\sim$0.4 km s$^{-1}$ and improve the signal-to-noise ratio.

**Table 2**
Observational Parameters of the KVN Four Bands

| Band | Representative Freq. (GHz) | HPBW (arcsec) | $\eta_A$ (%) | DPFU (K Jy$^{-1}$) |
|---|---|---|---|---|
| K | 22.235 | 125 | 61 | 0.076 |
| Q | 43.122 | 63 | 62 | 0.078 |
| W | 86.243 | 32 | 51 | 0.063 |
| D | 129.363 | 23 | 37 | 0.046 |

The system noise temperatures were consistently maintained within the ranges of 100–200 K (K), 100–300 K (Q), 180–500 K (W), and 200–700 K (D), depending on the weather conditions and elevation of the observation target. During the 12 yr observation period, the performance metrics of the KVN telescopes, i.e., the half-power beamwidth (HPBW), aperture efficiency ($\eta_A$), and degree per flux density unit (DPFU), were measured several times. The average values are presented in Table 2 for convenience, although we analyzed the actual observed data using antenna parameters corresponding to the observation period.[4]

To obtain the antenna temperature, $T_A^*$ (kelvin), the calibration was performed using the chopper wheel method to account for atmospheric attenuation and variations in antenna gain. Sky-dipping curve analysis was performed to correct for the atmospheric opacity. The integration time for achieving the 1$\sigma$ sensitivity (rms = 0.01–0.05 K) was 90–120 minutes in the position-switching mode. All observational data were analyzed using the GILDAS[5] software package. Subsequently, $T_A^*$ was transformed into the flux density using conversion factors (Jy K$^{-1}$) corresponding to the observation dates.

## 3. Results

Figure 1 presents simultaneously obtained spectra of the detected SiO43; SiO42; SiO86; SiO129; $^{29}$SiO $v=1$, $J=1-0$; and H$_2$O lines during the monitoring periods toward WX Ser. The red dashed line in each spectrum represents the stellar radial velocity ($V_*$) of 5.95 km s$^{-1}$, as determined from OH maser observations (J. Herman & H. J. Habing 1985). Colored intensity maps are shown in Figure 2, displaying the temporal variations of the SiO43, SiO42, SiO86, SiO129, and H$_2$O masers relative to the LSR velocity and optical phase. The color scales represent the detected intensity of each maser. The monitoring data were analyzed using techniques specified by A. Winnberg et al. (2008) and J. Brand et al. (2020), who confirmed the overall pattern through linear interpolation of the observed H$_2$O maser data for semiregular variable stars. The measured properties of SiO and H$_2$O masers are summarized in Tables 3 and 4. In these tables, $V_{\text{peak}}$ represents the velocities of the channels with the peak flux density identified in the individual line profiles, and $V_{\text{mean}}$ is the velocity centroid corresponding to the first moment of each line profile, which is the same parameter used in J. Kim et al. (2019) and references therein. The $^{28}$SiO $v=0$, $J=1-0$, $J=2-1$ and $v=2$, $J=2-1$, and $J=3-2$ lines were not detected at rms noise levels ranging from 0.1 to 0.5 Jy, while the emission of the $^{29}$SiO $v=0$, $J=1-0$ line was detected at three epochs. The $^{29}$SiO $v=0$, $J=1-0$ maser line had been predominantly detected with weak intensity, consistently displaying single-peaked profiles, and no observations had been conducted since 2012.

---
[4] https://radio.kasi.re.kr/status_report.php
[5] https://www.iram.fr/IRAMFR/GILDAS

Figures 1 and 2 show that, similar to other Mira variable stars, the strongest components on the SiO maser spectra prominently appear near the $V_*$, with sometimes several subcomponents. Table 3 demonstrates that the peak velocities of SiO masers varied within a range of ~3 km s$^{-1}$, randomly appearing in both the blue- and redshifted velocity regions relative to the $V_*$ for 12 yr. We have identified six peak features in SiO43 and SiO42 spectra, varying their relative intensities over the observed epochs. The six peaks were simultaneously detected at, e.g., epoch 16 (marked with the downward-pointing arrows at epoch 16 in Figure 1); however, at most of the epochs many of the six peaks were undetected. Notice that these six peak features appear in between −1 and 13 km s$^{-1}$ (±7 km s$^{-1}$ with respect to the $V_*$). The velocity distribution patterns of these two masers can be readily seen in Figures 2(a) and (b). These masers seem to appear and disappear with redshift and/or blueshift relative to the $V_*$ during our monitoring periods without any regular pattern. This indicates most likely the presence of turbulent material motion in the SiO masing region. The intensities of the SiO86 maser were considerably lower than those of SiO43 and SiO42 masers, and five peaks were identified in the spectral range of −1.4 km s$^{-1}$ < $V_{\text{LSR}}$ < 12.2 km s$^{-1}$ during the observed epochs. They are marked with the downward-pointing arrows at epochs 15 and 16 in Figure 1. In Figure 2(c), the components on SiO86 maser spectra exhibited a less dispersed shape compared to SiO43 and SiO42, since its maser intensities are slightly weaker than them. The intensity of detected SiO129 is noticeably (more than three times) weaker than the other SiO masers, and the number of detected components was also lower (at most four peaks are found, marked with the downward-pointing arrows at epoch 15 in Figure 1). Out of a total of 36 observed epochs, SiO129 was even not detected at all in four instances. Thus, in Figure 2(d), it is hard to identify the turbulent flow in the region, as the components mainly appear near the $V_*$.

We detected six components in the H$_2$O maser emission, as shown in Table 4. These components were categorized relative to the $V_*$; four peaks were identified as the blueshifted line labeled from B1 to B4, while two peaks were identified as the redshifted components labeled with R1 and R2 (i.e., B4 component at around $V_{\text{LSR}} = -2$ km s$^{-1}$, B3 component at around −1 km s$^{-1}$, B2 component at around +1.5 km s$^{-1}$, B1 component at around +4 km s$^{-1}$ on the epoch 9 ($\phi = 2.44$) spectrum, R1 component at around +7 km s$^{-1}$, and R2 component at around +8 km s$^{-1}$ on epoch 8 ($\phi = 2.22$) spectrum, in Figure 1). Among the six components, B4, B2, R1, and R2 were detected in most of the epochs, while B3 and B1 were only detected seven and six times, respectively. A very interesting thing that is shown in Figure 2(e) is that, unlike SiO masers, the velocity distribution of the H$_2$O maser gradually moves away from the $V_*$, suggesting an outward acceleration. Upon detailed analysis of the six components, it is noted that the B4, B2, and B1 components show a tendency to drift away from the $V_*$ at about 0.5–1.3 km s$^{-1}$ for 12 yr. In addition, the R1 and R2 components also show a gradual departure from the $V_*$ at about 0.4–1.0 km s$^{-1}$, although their divergence is less pronounced compared to the peaks identified in the blueshifted region. Based on the velocity structure identified, it is anticipated that the CSE of WX Ser may have an asymmetric shell structure, potentially accounting for the intrinsic characteristics of the asymmetric material outflow. However, to achieve a clear understanding of this phenomenon, it is necessary to conduct simultaneous very long baseline interferometry (VLBI) observations of SiO and H$_2$O masers, which can pinpoint the location of the central star astrometrically (e.g., R. Dodson et al. 2018; D.-J. Kim et al. 2018; D.-H. Yoon et al. 2018).

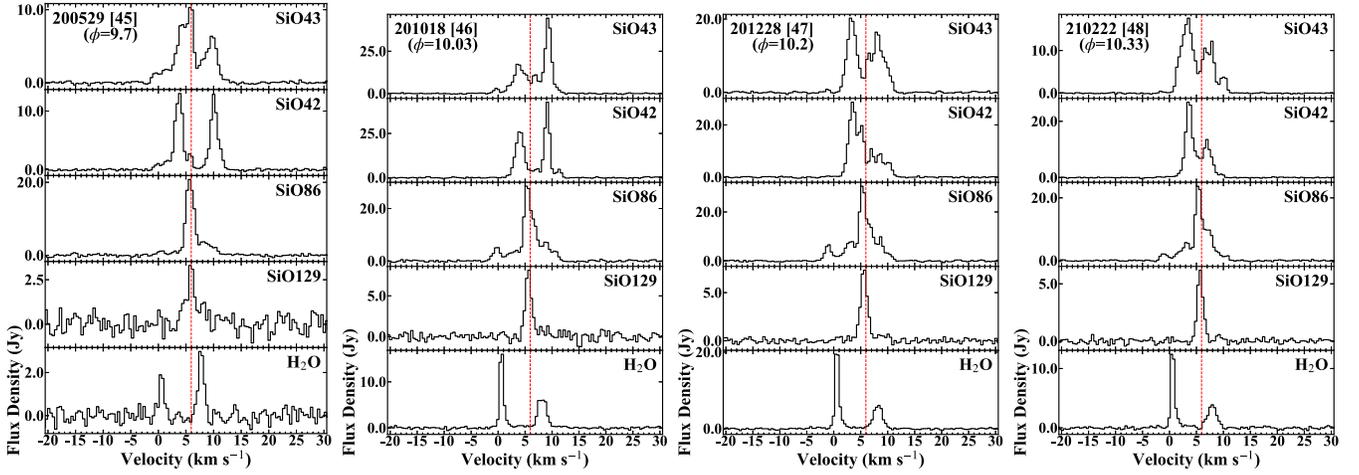

**Figure 1.** The spectra of SiO and $H_2O$ maser lines in 50 epochs. The intensity is expressed in terms of the flux density (Jy), while the x-axis represents the LSR velocity, $V_{LSR}$ (kilometers per second). The vertical red dashed lines indicate the stellar radial velocity ($V_* = 5.95$ km s$^{-1}$). The date of observation (in the yymmdd format) is displayed in the upper left corner of each spectrum, along with the observation epoch and optical phase ($\phi$). Four example spectra are shown with SiO43, SiO42, SiO86, SiO129, and $H_2O$ spectra.
(The complete figure set (50 images) is available in the online article.)

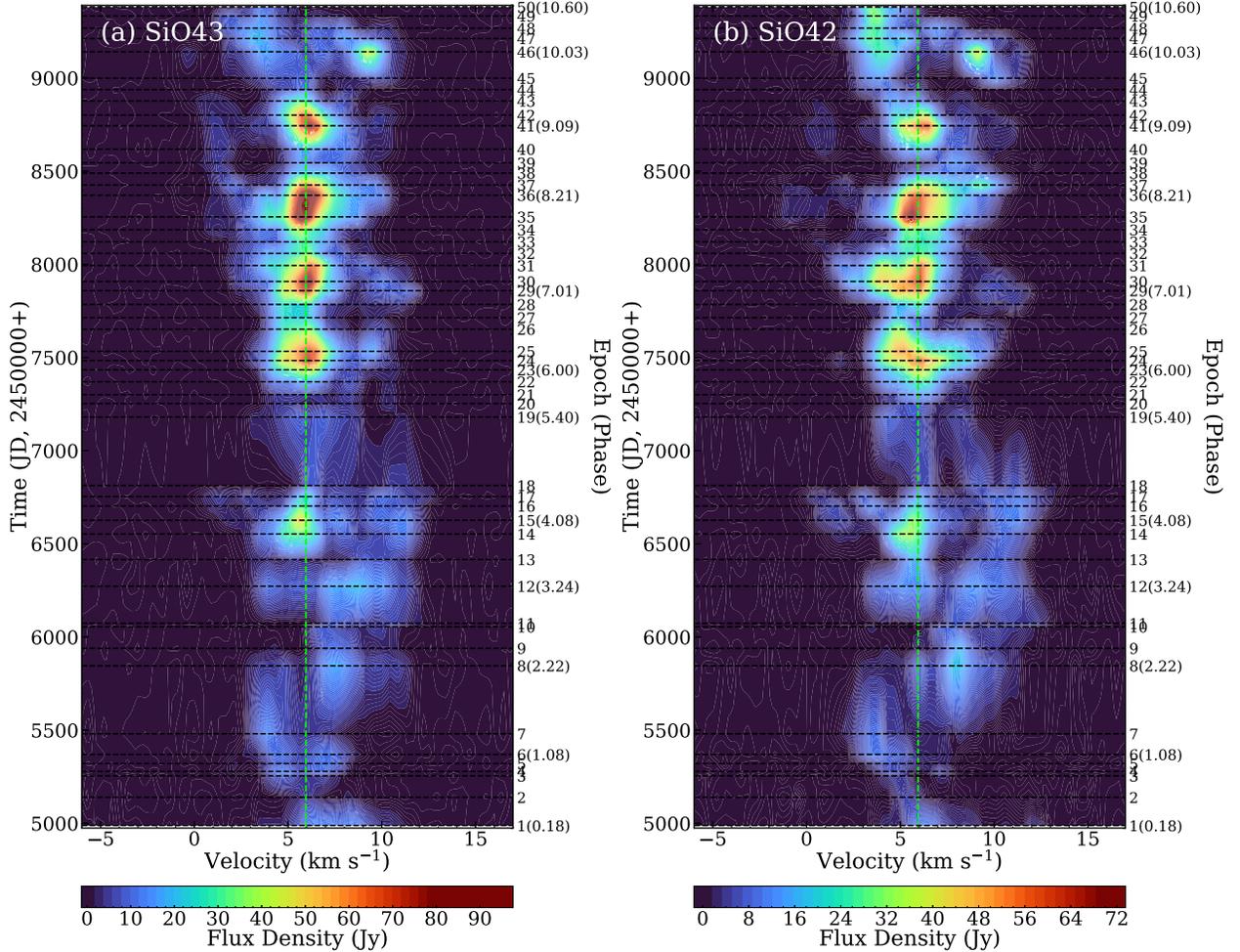

**Figure 2.** Color intensity maps of (a) SiO43 maser over 50 epochs, (b) SiO42 maser over 50 epochs, (c) SiO86 maser over 41 epochs, (d) SiO129 maser over 36 epochs, and (e) $H_2O$ maser over 50 epochs. (f) Enlarged spectra of the $H_2O$ maser line for specific dates ($\phi = 8.79$–9.41) to show a flare epoch 41 ($\phi = 9.09$). The blue- and redshifted components are annotated with arrows. Each color map shows time variations of monitored SiO and $H_2O$ masers based on the LSR velocity and optical phase. The horizontal dashed lines represent observed epochs, and the vertical green dashed lines represent the $V_*$. The left side of the y-axis represents observed times (JD+2450000), while the right-side axis represents the observed epochs, along with the optical phases in parentheses. The flux density scales of each maser are expressed by the color bar at the bottom.

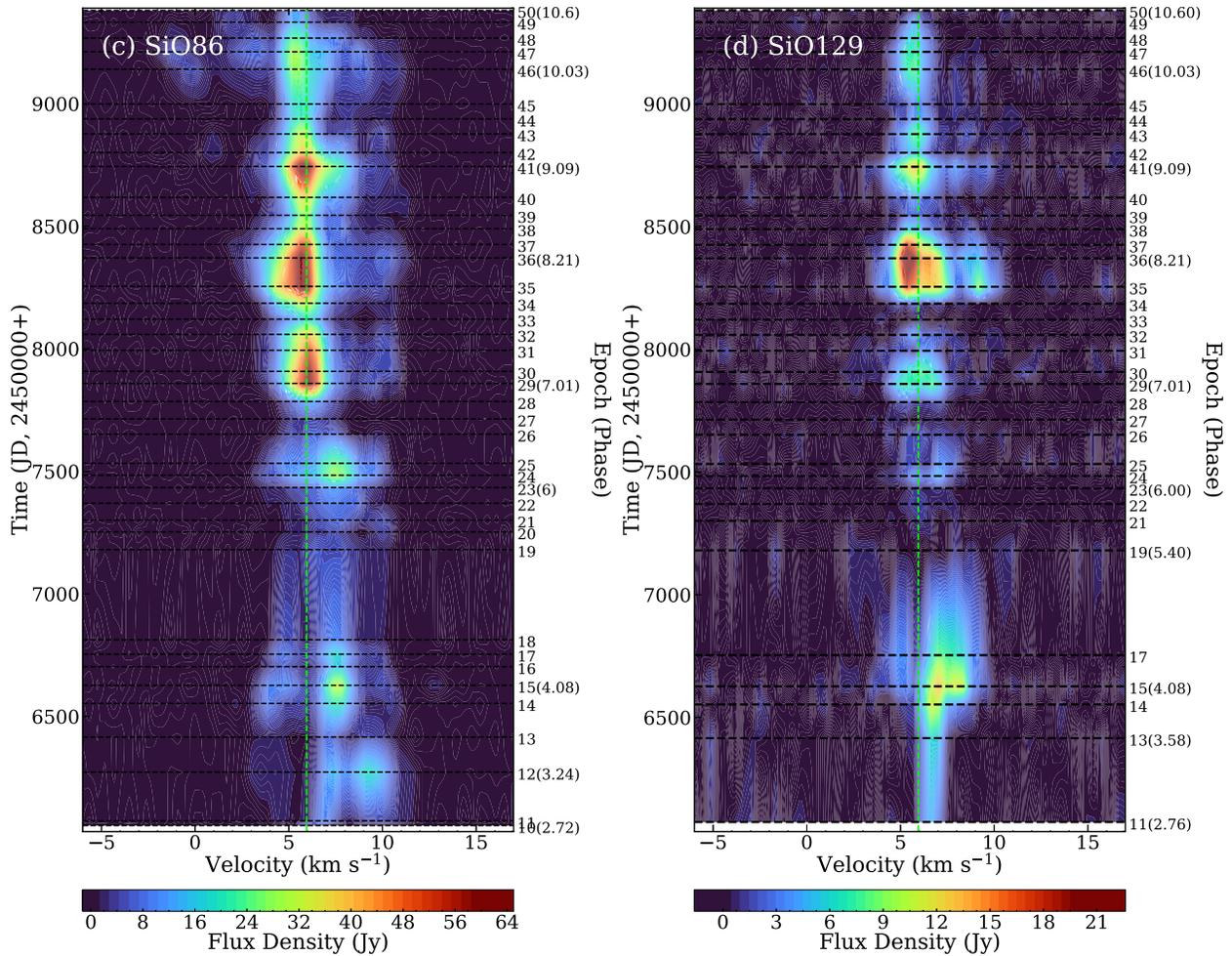

**Figure 2.** (Continued.)

The flux densities of the H$_2$O masers show greater variability compared to the SiO masers, including variations in their line profiles. In particular, the flux densities in the blueshifted components demonstrate more temporal variations than the redshifted components, suggesting a high level of activity. Figure 2(f) provides a close-up view of the temporal behavior of a flaring H$_2$O maser, which exhibits an obvious bias toward blueshifted velocity regions. On epoch 40 ($\phi = 8.79$), the H$_2$O maser was detected as only two components, namely, B2 and R1 components, with intensities of 8.4 and 1.9 Jy at velocities of ∼0.8 and ∼7.6 km s$^{-1}$, respectively. However, approximately 4 months later, on epoch 41 ($\phi = 9.09$), the H$_2$O maser exhibited the presence of four distinct components, i.e., the B4, B2, R1, and R2 components. Notably, the B4 component displays a significant intensity of 41.5 Jy at a velocity of −3.0 km s$^{-1}$. The flux density of the B2 component is approximately 14 times greater (120.8 Jy at ∼0.8 km s$^{-1}$) than that on epoch 40. On the other hand, the flux densities of the R1 and R2 components appear to be relatively stable, with changes ranging from ∼2 to ∼7 Jy. After the event, only the B2 and R1 components appear to have survived in the H$_2$O maser region, and the flux density of the B2 component indicates weak emissions, with values of 16.0 Jy at epoch 46 ($\phi = 10.03$) and 19.5 Jy at epoch 47 ($\phi = 10.20$). G. M. Rudnitskij (2008) suggested that the transient amplification occasionally detected in H$_2$O masers could be caused by flare events occurring on the stellar surface, a phenomenon frequently observed in Mira variable stars.

A comprehensive analysis and discussion of the maser properties are presented in Section 4.

## 4. Analyses and Discussion

### 4.1. Optical Variability

The first photometric study of WX Ser at the optical wavelength was published by F. E. Ross (1927). Subsequently, W. P. Bidelman (1954) reported that the pulsation period of WX Ser is 425 days, based on data provided by AAVSO. This period was confirmed by J. Herman & H. J. Habing (1985) through a 2 yr monitoring of the 1612 MHz OH maser. S. D. Price et al. (2010) also confirmed the period from the 3 yr near-IR observation data using COBE/DIRBE.

The OLC of a Mira variable star can typically be represented as a sinusoidal waveform. The OLC of WX Ser has been confirmed to closely follow a sinusoidal curve, based on the optical data provided by AAVSO. We applied a tilted sinusoidal curve, $y(t) = \sin(t + \frac{1}{2}\sin(t))$, which is a slight modification of the sinusoidal curve, to fit the OLC of WX Ser and determined the day of maximum brightness during our monitoring period (see Figure 3). During the monitoring period of WX Ser from 2009 to 2021, which spans approximately 10 pulsating optical cycles, the initial five cycles lacked sufficient data, necessitating the use of the same coefficients for the fitting process. Subsequently, for the sixth through 10th cycles, where sufficient data points were available, individualized curve fitting was conducted to capture the pulsation period of each cycle. Based on the fitting analysis, we determined a pulsation period of $421 \pm 4$ days, leading to the following ephemeris:

$$\begin{aligned}\text{Max.} = \text{JD } 2450000 &+ 5335(1\text{st}), 5756(2\text{nd}), \\ &\times 6173(3\text{rd}), 6594(4\text{th}), 7015(5\text{th}), 7434 \text{ (6th)}, \\ &\times 7856(7\text{th}), 8281(8\text{th}), 8706(9\text{th}), 9127(10\text{th}).\end{aligned}$$
(1)

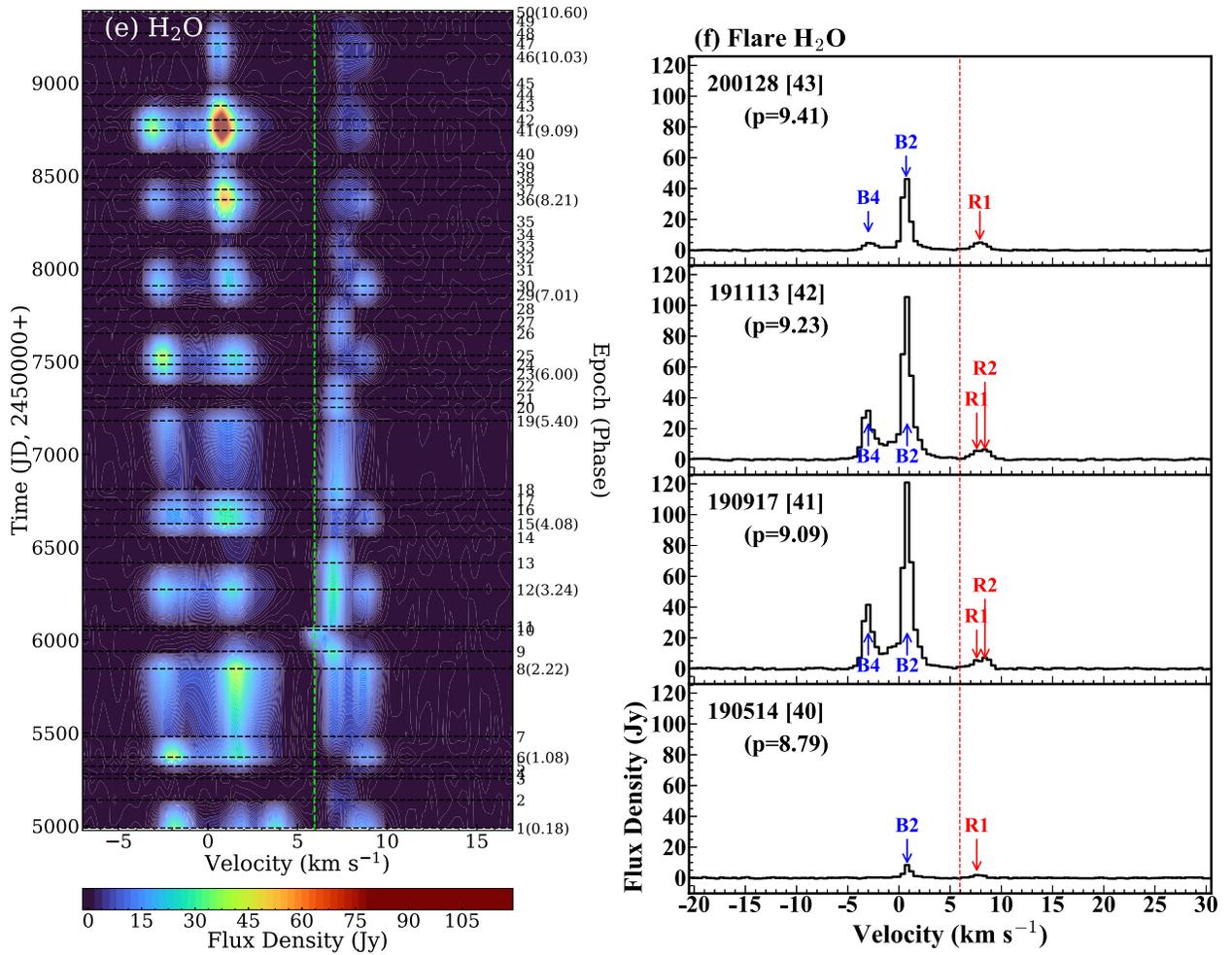

**Figure 2.** (Continued.)

Our estimated pulsation period is consistent with the value reported by AAVSO within the uncertainty limit. Therefore, we conclude that WX Ser is currently exhibiting a stable pulsation cycle with a consistent period, comparable to the period observed 60 yr ago (~425 days).

### 4.2. Time Variations of SiO and $H_2O$ Maser Fluxes

We analyzed the temporal variations of the SiO43, SiO42, SiO86, SiO129, and $H_2O$ maser lines in comparison with the OLC over the same period.

Figure 4 shows the temporal variations in peak flux densities of these maser lines, with the guidelines (vertical dotted lines) indicating the maximum phases of the OLC. The color scales of each maser spectrum were determined by the peak velocities of the detected components. Figure 5 displays the temporal variations in the integrated flux densities of the masers, exhibiting a similar pattern to that of Figure 4. To compare the variation in maser intensities with the optical maximum phases, we applied polynomial interpolation to the observation points in Figure 5. Here, in order to avoid overestimation, we did not interpolate for periods with two interruptions (i.e., from epoch 7 to epoch 8 and from epoch 18 to epoch 19) and for the 7 month gap between epoch 11 ($\phi = 2.76$) and epoch 12 ($\phi = 3.24$).

#### 4.2.1. Characteristics of Four SiO Maser Lines

As shown in Figures 4 and 5, the intensity variations of all SiO masers are well aligned with the OLC variation pattern. It should be noted, however, that the amplitudes of the maser curves vary across cycles. Since 2014, the amplitudes of all SiO masers have been increasing until the eighth optical maximum in 2018, following which they have begun to decrease. This trend underscores the dynamic nature of the SiO maser behavior in response to the evolving pulsation dynamics of the central star.

Table 3
Observational Results of SiO Transitions

| Maser Line | Obs. Epoch | Date (Phase) (yymmdd) | $S_{\rm peak}$ (Jy) | $\int S_\nu\, dv$ (Jy km s$^{-1}$) | rms (Jy) | $V_{\rm peak}$ (km s$^{-1}$) | $V_{\rm mean}$ (km s$^{-1}$) |
|---|---|---|---|---|---|---|---|
| SiO43 | 1 | 090605 (0.18) | 17.4, 17.4 | 67.0 | 0.457 | 6.5, 8.9 | 7.4 |
| | 2 | 091104 (0.54) | 5.6, 2.5 | 10.4 | 0.307 | 6.0, 8.7 | 6.6 |
| | 3 | 100227 (0.81) | 9.4, 6.4 | 26.7 | 0.301 | 4.7, 6.9 | 5.6 |
| | 4 | 100323 (0.87) | 2.0, 13.5, 11.9 | 43.9 | 0.283 | 2.1, 4.7, 6.9 | 5.6 |
| | 5 | 100504 (0.97) | 2.5, 9.8, 12.4 | 45.3 | 0.278 | 2.1, 4.7, 6.9 | 5.7 |
| | 6 | 100622 (1.08) | 11.5, 12.0, 1.8 | 54.5 | 0.292 | 4.7, 7.1 10.1 | 6.0 |
| | 7 | 101012 (1.35) | 14.6, 5.4 | 33.5 | 0.364 | 3.9, 6.5 | 4.9 |
| | 8 | 111012 (2.22) | 7.1, 17.7, 6.5, 5.5 | 66.6 | 0.264 | 4.8, 7.4, 9.5, 10.1 | 7.5 |
| | 9 | 120103 (2.44) | 1.5, 2.4, 11.8, 3.8, 1.3 | 38.9 | 0.252 | 3.0, 4.3, 7.4, 10.4, 12.6 | 7.8 |
| | 10 | 120507 (2.72) | 1.8, 9.1, 5.2, 1.0 | 36.6 | 0.213 | 3.5, 7.0, 9.6, 12.6 | 7.9 |
| | 11 | 120526 (2.76) | 1.0, 1.4, 9.5, 6.4, 6.0 | 37.1 | 0.222 | 3.8, 4.7, 7.8, 9.1, 10.4 | 8.7 |
| | 12 | 121211 (3.24) | 11.4, 9.8, 18.1, 21.2, 15.6 | 107.7 | 0.389 | 3.9, 6.2, 7.4, 8.7, 10.0 | 7.7 |
| | 13 | 130503 (3.58) | 3.6, 4.3, 9.0, 7.7 | 39.6 | 0.238 | 3.9, 5.2, 9.1, 10.0 | 7.9 |
| | 14 | 130917 (3.90) | 32.2, 9.7, 12.3 | 129.6 | 0.193 | 5.2, 7.8, 10.9 | 6.7 |
| | 15 | 131130 (4.08) | 3.1, 52.9, 14.4, 11.5, 14.0 | 148.3 | 0.168 | 2.0, 5.7, 7.8, 10.0, 11.0 | 6.8 |
| | 16 | 140214 (4.26) | 4.3, 5.9, 31.3, 9.7, 10.2 | 98.8 | 0.153 | 1.3, 3.0, 5.6, 7.4, 10.0 | 6.8 |
| | 17 | 140406 (4.38) | 5.1, 8.3, 25.9, 7.7, 10.5 | 91.8 | 0.179 | 1.3, 2.6, 6.1, 7.4, 10.0 | 6.9 |
| | 18 | 140604 (4.52) | 0.6, 0.5, 8.5, 4.6, 4.9 | 32.2 | 0.185 | 1.7, 3.5, 6.1, 7.8, 10.5 | 8.6 |
| | 19 | 150607 (5.40) | 9.8, 9.8, 2.4 | 44.7 | 0.254 | 5.6, 7.0, 11.3 | 6.6 |
| | 20 | 150818 (5.57) | 4.4, 4.3, 4.5, 3.1 | 25.1 | 0.205 | 4.8, 7.0, 8.3, 10.5 | 7.1 |
| | 21 | 151005 (5.68) | 5.7, 5.8, 3.2 | 29.2 | 0.149 | 5.2, 8.3, 10.4 | 7.0 |
| | 22 | 151213 (5.85) | 18.1, 8.3, 2.6 | 61.7 | 0.142 | 5.6, 8.3, 10.4 | 6.3 |
| | 23 | 160215 (6.00) | 46.1, 7.2, 3.1 | 129.1 | 0.163 | 6.1, 8.2, 10.5 | 6.0 |
| | 24 | 160405 (6.12) | 75.4, 11.2 | 207.8 | 0.133 | 6.1, 9.1 | 6.1 |
| | 25 | 160524 (6.23) | 68.7, 18.5 | 215.1 | 0.199 | 6.1, 9.5 | 6.3 |
| | 26 | 160920 (6.52) | 37.4, 6.4 | 99.5 | 0.167 | 6.1, 9.6 | 6.1 |
| | 27 | 161121 (6.66) | 25.2, 22.1, 3.0 | 65.5 | 0.142 | 4.8, 6.1, 9.6 | 5.7 |
| | 28 | 170131 (6.83) | 27.2, 24.5, 3.8, 4.8 | 83.5 | 0.139 | 4.8, 6.1, 9.6, 10.9 | 6.0 |
| | 29 | 170415 (7.01) | 70.7, 10.2 | 194.8 | 0.212 | 6.1, 11.1 | 6.4 |
| | 30 | 170603 (7.12) | 90.6, 12.8 | 213.8 | 0.132 | 6.1, 10.0 | 6.2 |
| | 31 | 170828 (7.32) | 65.2, 10.2, 3.7 | 192.1 | 0.221 | 6.1, 9.2, 10.5 | 6.0 |
| | 32 | 171101 (7.48) | 17.2, 36.4, 9.5 | 106.6 | 0.130 | 3.9, 6.1, 7.9 | 5.5 |
| | 33 | 180102 (7.62) | 9.7, 31.4, 2.7 | 92.1 | 0.207 | 3.5, 5.7, 9.2 | 5.5 |
| | 34 | 180308 (7.78) | 3.7, 6.3, 40.2 | 116.6 | 0.139 | 1.4, 3.5, 5.7 | 5.4 |
| | 35 | 180516 (7.94) | 2.7, 97.5 | 245.7 | 0.230 | 0.9, 5.7 | 5.8 |
| | 36 | 180909 (8.21) | 18.7, 98.8 | 270.7 | 0.175 | 4.1, 6.1 | 6.2 |
| | 37 | 181103 (8.34) | 10.6, 57.9, 13.6 | 181.7 | 0.227 | 2.2, 6.1, 9.2 | 6.1 |
| | 38 | 190106 (8.49) | 5.5, 21.8 | 79.9 | 0.163 | 1.3, 5.7 | 6.0 |
| | 39 | 190302 (8.62) | 4.8, 0.6, 15.9, 11.5 | 53.5 | 0.141 | 1.3, 3.5, 5.7, 7.5 | 6.1 |
| | 40 | 190514 (8.79) | 5.0, 1.4, 17.4, 12.7, 3.0 | 65.7 | 0.243 | 1.3, 3.1, 6.1, 7.9, 10.0 | 6.2 |
| | 41 | 190917 (9.09) | 4.5, 84.9, 4.6 | 191.0 | 0.196 | 0.9, 6.1, 9.6 | 6.1 |
| | 42 | 191113 (9.23) | 4.3, 64.6, 4.5 | 161.9 | 0.214 | 0.9, 6.1, 10.0 | 5.9 |
| | 43 | 200128 (9.41) | 3.6, 32.0, 8.2 | 77.5 | 0.173 | 0.9, 5.7, 7.9 | 5.4 |
| | 44 | 200329 (9.55) | 1.7, 1.4, 10.9, 4.5 | 31.1 | 0.202 | −0.3, 0.9, 5.7, 8.3 | 5.3 |
| | 45 | 200529 (9.70) | 1.4, 1.9, 8.1, 10.4, 6.5 | 54.1 | 0.134 | −0.4, 1.3, 4.4, 5.7, 9.6 | 6.0 |
| | 46 | 201018 (10.03) | 3.0, 17.1, 11.1, 44.4 | 133.9 | 0.187 | −0.4, 3.5, 6.8, 9.2 | 6.9 |
| | 47 | 201228 (10.20) | 0.9, 20.1, 10.8, 16.5 | 100.5 | 0.182 | −1.3, 3.1, 6.5, 7.9 | 5.8 |
| | 48 | 210222 (10.33) | 0.5, 17.6, 10.5, 12.1, 3.7 | 82.2 | 0.101 | −1.6, 3.5, 6.6, 7.9, 10.0 | 5.0 |
| | 49 | 210428 (10.49) | 9.4, 14.4, 3.3, 2.4 | 41.6 | 0.177 | 3.5, 6.6, 7.9 | 3.8 |
| | 50 | 210615 (10.60) | 0.8, 11.0, 1.5 | 27.6 | 0.178 | 0.5, 3.5, 7.0 | 4.0 |
| SiO42 | 1 | 090605 (0.18) | 2.7, 11.7, 4.0, 3.6 | 37.1 | 0.439 | 4.6, 6.3, 8.5, 12.0 | 7.3 |
| | 2 | 091104 (0.54) | 5.8, 1.3 | 11.0 | 0.317 | 5.3, 7.9 | 5.9 |
| | 3 | 100227 (0.81) | 5.0, 2.1 | 12.3 | 0.340 | 4.8, 7.4 | 5.3 |
| | 4 | 100323 (0.87) | 7.1, 5.5 | 18.5 | 0.308 | 4.4, 7.0 | 5.5 |
| | 5 | 100504 (0.97) | 7.1, 2.4 | 20.3 | 0.274 | 4.4, 7.1 | 5.0 |
| | 6 | 100622 (1.08) | 10.9 | 31.3 | 0.287 | 3.9 | 4.6 |
| | 7 | 101012 (1.35) | 10.0, 2.9, 2.7 | 28.8 | 0.370 | 3.8, 6.9, 8.2 | 4.6 |
| | 8 | 111012 (2.22) | 3.1, 9.8, 19.3 | 62.2 | 0.275 | 4.2, 6.0, 7.8 | 7.7 |
| | 9 | 120103 (2.44) | 1.7, 3.4, 18.1, 3.6, 1.5 | 40.0 | 0.274 | 4.3, 6.0, 8.2, 10.0, 12.1 | 8.2 |
| | 10 | 120507 (2.72) | 1.9, 1.7, 8.8, 5.8, 3.7 | 34.4 | 0.215 | 3.4, 5.1, 7.4, 10.0, 11.3 | 8.3 |
| | 11 | 120526 (2.76) | 1.7, 2.7, 7.9, 5.6, 5.5 | 38.6 | 0.255 | 4.2, 6.0, 8.2, 10.0, 11.7 | 9.1 |
| | 12 | 121211 (3.24) | 13.5, 14.2 | 87.3 | 0.346 | 5.2, 9.1 | 7.4 |
| | 13 | 130503 (3.58) | 16.7, 7.9 | 43.7 | 0.265 | 5.6, 10.0 | 7.7 |
| | 14 | 130917 (3.90) | 2.6, 34.6, 4.5, 9.6 | 101.7 | 0.215 | 1.7, 5.2, 7.8, 10.4 | 6.4 |
| | 15 | 131130 (4.08) | 5.7, 33.3, 11.9 | 110.9 | 0.186 | 1.7, 5.6, 10.9 | 6.6 |
| | 16 | 140214 (4.26) | 6.5, 12.4, 20.7, 6.1, 5.7, 11.8 | 99.2 | 0.167 | 1.3, 3.5, 5.7, 7.4, 9.2, 11.3 | 6.8 |
| | 17 | 140406 (4.38) | 4.0, 8.6, 15.7, 9.0, 10.5 | 76.0 | 0.187 | 0.9, 3.0, 6.1, 10.0, 11.4 | 7.3 |
| | 18 | 140604 (4.52) | 0.7, 6.5, 7.6, 6.5, 1.7 | 32.7 | 0.171 | 1.7, 6.5, 9.1, 10.9, 13.0 | 9.0 |
| | 19 | 150607 (5.40) | 9.4, 7.3, 5.5, 0.9 | 32.8 | 0.240 | 4.3, 5.6, 8.2, 11.3 | 6.0 |
| | 20 | 150818 (5.57) | 7.7, 2.8 | 23.7 | 0.211 | 6.0, 7.9 | 6.0 |

Table 3
(Continued)

| Maser Line | Obs. Epoch | Date (Phase) (yymmdd) | $S_{\rm peak}$ (Jy) | $\int S_\nu\, dv$ (Jy km s$^{-1}$) | rms (Jy) | $V_{\rm peak}$ (km s$^{-1}$) | $V_{\rm mean}$ (km s$^{-1}$) |
|---|---|---|---|---|---|---|---|
| | 21 | 151005 (5.68) | 13.4, 2.7 | 36.2 | 0.147 | 6.1, 8.3 | 6.0 |
| | 22 | 151213 (5.85) | 25.2 | 73.4 | 0.151 | 6.5 | 5.9 |
| | 23 | 160215 (6.00) | 1.4, 43.9 | 125.9 | 0.173 | 1.9, 6.1 | 6.0 |
| | 24 | 160405 (6.12) | 2.9, 58.9 | 211.7 | 0.150 | 1.7, 6.1 | 6.1 |
| | 25 | 160524 (6.23) | 52.0, 24.9 | 182.2 | 0.197 | 5.2, 7.4 | 6.1 |
| | 26 | 160920 (6.52) | 33.7, 4.7, 7.5, 3.8 | 75.0 | 0.174 | 4.8, 7.8, 8.7, 10.0 | 5.8 |
| | 27 | 161121 (6.66) | 17.8, 10.0, 1.5 | 38.7 | 0.133 | 4.8, 6.1, 9.1 | 5.7 |
| | 28 | 170131 (6.83) | 23.1, 2.7, 3.8, 3.8 | 60.5 | 0.130 | 6.1, 8.3, 9.6, 10.9 | 6.3 |
| | 29 | 170415 (7.01) | 49.2, 54.6, 6.2 | 177.1 | 0.211 | 4.6, 6.1, 9.5 | 5.6 |
| | 30 | 170603 (7.12) | 19.2, 46.6, 62.3, 5.3 | 217.6 | 0.150 | 2.5, 3.9, 6.1, 10.0 | 5.2 |
| | 31 | 170828 (7.32) | 11.5, 36.5, 58.9, 3.8 | 180.4 | 0.204 | 2.1, 4.3, 6.1, 9.2 | 5.3 |
| | 32 | 171101 (7.48) | 4.9, 22.5, 27.0 | 96.1 | 0.146 | 1.7, 5.2, 6.1 | 5.4 |
| | 33 | 180102 (7.62) | 2.3, 27.2 | 87.2 | 0.211 | 0.8, 5.2 | 5.7 |
| | 34 | 180308 (7.78) | 0.9, 2.8, 30.5, 28.5 | 82.4 | 0.129 | −0.1, 3.4, 5.6, 6.9 | 6.0 |
| | 35 | 180516 (7.94) | 3.0, 2.2, 74.5 | 205.2 | 0.231 | −0.9, 0.8, 5.6 | 5.9 |
| | 36 | 180909 (8.21) | 2.8, 3.3, 6.9, 58.5 | 193.6 | 0.185 | −0.5, 2.1, 3.5, 6.0 | 6.2 |
| | 37 | 181103 (8.34) | 1.9, 4.0, 2.2, 39.2, 29.4 | 154.2 | 0.211 | −0.9, 2.1, 3.5, 6.1, 9.1 | 7.0 |
| | 38 | 190106 (8.49) | 1.7, 11.6, 9.0 | 51.8 | 0.163 | −0.1, 6.9, 9.1 | 7.1 |
| | 39 | 190302 (8.62) | 3.4, 13.4, 1.2 | 35.6 | 0.182 | 5.2, 7.8, 12.2 | 7.2 |
| | 40 | 190514 (8.79) | 2.8, 6.0, 15.5, 9.1, 2.1 | 61.4 | 0.228 | 0.8, 3.0, 6.1, 7.8, 10.0 | 6.2 |
| | 41 | 190917 (9.09) | 3.0, 59.9, 7.0, 3.7 | 159.4 | 0.205 | 0.8, 6.5, 8.3, 10.0 | 6.0 |
| | 42 | 191113 (9.23) | 2.5, 21.6, 37.6, 4.2 | 86.3 | 0.201 | 0.8, 4.8, 6.1, 8.2 | 5.9 |
| | 43 | 200128 (9.41) | 1.8, 14.3, 12.8, 3.4, 1.3 | 44.6 | 0.168 | 0.4, 4.8, 6.1, 8.3, 11.3 | 5.4 |
| | 44 | 200329 (9.55) | 1.4, 9.4, 1.6, 1.1 | 33.6 | 0.211 | 1.3, 4.3, 8.3, 11.3 | 4.8 |
| | 45 | 200529 (9.70) | 1.0, 1.6, 13.0, 2.7, 12.9 | 49.2 | 0.125 | −0.1, 1.7, 3.9, 5.6, 10.0 | 6.4 |
| | 46 | 201018 (10.03) | 25.9, 5.0, 42.5, 4.7 | 115.5 | 0.216 | 3.9, 7.4, 9.1, 11.3 | 6.7 |
| | 47 | 201228 (10.20) | 28.7, 19.8, 10.8, 9.0, 5.7 | 107.7 | 0.181 | 3.4, 5.2, 6.9, 8.7, 10.1 | 5.3 |
| | 48 | 210222 (10.33) | 26.8, 13.5, 1.3 | 78.7 | 0.110 | 3.5, 6.9, 9.6 | 4.9 |
| | 49 | 210428 (10.49) | 31.9, 1.5 | 59.6 | 0.178 | 3.9, 6.9 | 4.1 |
| | 50 | 210615 (10.60) | 1.6, 32.7 | 59.3 | 0.169 | 1.3, 3.4 | 3.9 |
| SiO86 | 10 | 120507 (2.72) | 7.5, 1.5 | 13.7 | 0.289 | 6.1, 10.2 | 7.4 |
| | 11 | 120526 (2.76) | 2.0, 10.8, 4.9 | 29.8 | 0.317 | 4.3, 6.9, 10.1 | 8.0 |
| | 12 | 121211 (3.24) | 2.4, 15.0, 19.5 | 66.5 | 0.148 | 3.5, 7.4, 9.1 | 8.3 |
| | 13 | 130503 (3.58) | 1.8, 7.8, 4.9 | 22.7 | 0.300 | 4.3, 6.9, 9.1 | 7.5 |
| | 14 | 130917 (3.90) | 1.8, 9.4, 4.8, 20.2, 3.6 | 59.9 | 0.310 | 2.2, 3.8, 5.7, 7.4, 10.4 | 6.7 |
| | 15 | 131130 (4.08) | 1.6, 10.9, 31.4, 5.2, 2.0 | 87.4 | 0.232 | 1.7, 4.7, 7.8, 10.0, 13.0 | 7.0 |
| | 16 | 140214 (4.26) | 1.6, 6.5, 15.4, 4.8, 7.5 | 50.4 | 0.218 | 1.3, 4.8, 7.4, 10.0, 12.2 | 7.0 |
| | 17 | 140406 (4.38) | 0.9, 9.0, 18.8, 4.8 | 56.1 | 0.228 | 2.2, 4.8, 7.4, 9.6 | 7.0 |
| | 18 | 140604 (4.52) | 5.7, 10.3, 2.3 | 31.9 | 0.371 | 4.8, 7.4, 9.5 | 7.0 |
| | 19 | 150607 (5.40) | 1.6, 3.2, 6.4, 5.8, 2.3 | 28.7 | 0.417 | 3.0, 4.8, 6.5, 7.8, 10.0 | 6.9 |
| | 20 | 150818 (5.57) | 1.5, 5.4, 4.1 | 19.1 | 0.392 | 3.4, 7.4, 10.0 | 7.5 |
| | 21 | 151005 (5.68) | 2.2, 7.3, 5.1 | 25.4 | 0.282 | 5.2, 7.8, 10.0 | 7.9 |
| | 22 | 151213 (5.85) | 1.7, 9.5 | 33.2 | 0.167 | 3.9, 7.4 | 7.2 |
| | 23 | 160215 (6.00) | 12.4 | 40.7 | 0.190 | 7.4 | 7.1 |
| | 24 | 160405 (6.12) | 29.5 | 99.5 | 0.213 | 7.4 | 7.1 |
| | 25 | 160524 (6.23) | 10.8, 25.0, 9.6 | 86.9 | 0.313 | 5.2, 7.4, 9.5 | 7.1 |
| | 26 | 160920 (6.52) | 0.7, 5.3, 7.1, 7.4, 2.0 | 25.9 | 0.276 | 3.0, 5.2, 6.5, 7.4, 9.5 | 6.9 |
| | 27 | 161121 (6.66) | 4.5 | 12.1 | 0.168 | 5.6 | 6.2 |
| | 28 | 170131 (6.83) | 12.9, 1.7 | 29.6 | 0.173 | 6.1, 9.5 | 6.4 |
| | 29 | 170415 (7.01) | 62.9, 6.3, 3.2 | 131.5 | 0.276 | 2.2, 6.1, 9.5, 10.4 | 6.2 |
| | 30 | 170603 (7.12) | 60.7, 5.9, 5.6 | 129.4 | 0.207 | 6.1, 9.5, 10.4 | 6.3 |
| | 31 | 170828 (7.32) | 51.7, 3.9, 5.5 | 108.6 | 0.432 | 6.1, 9.1, 10.0 | 6.3 |
| | 32 | 171101 (7.48) | 39.9, 3.9, 5.0 | 89.2 | 0.296 | 6.1, 8.2, 10.1 | 6.2 |
| | 33 | 180102 (7.62) | 17.5, 2.5 | 40.0 | 0.229 | 6.1, 10.4 | 6.2 |
| | 34 | 180308 (7.78) | 33.0, 8.6 | 86.2 | 0.279 | 6.1, 10.2 | 5.9 |
| | 35 | 180516 (7.94) | 64.7, 5.7 | 170.7 | 0.403 | 5.6, 9.5 | 5.8 |
| | 36 | 180909 (8.21) | 63.5, 7.8 | 165.1 | 0.220 | 5.6, 10.1 | 5.9 |
| | 37 | 181103 (8.34) | 52.4, 7.8 | 128.4 | 0.209 | 5.6, 10.0 | 5.9 |
| | 38 | 190106 (8.49) | 32.9, 9.5, 2.7 | 81.7 | 0.173 | 5.6, 7.4, 10.0 | 6.1 |
| | 39 | 190302 (8.62) | 27.9, 7.6, 2.1, 1.7 | 62.0 | 0.201 | 5.6, 7.4, 9.5, 11.3 | 6.3 |
| | 40 | 190514 (8.79) | 1.8, 35.8, 3.2 | 89.8 | 0.311 | 3.0, 6.1, 9.5 | 6.4 |
| | 41 | 190917 (9.09) | 65.6 | 176.3 | 0.333 | 5.6 | 6.2 |
| | 42 | 191113 (9.23) | 2.0, 38.1, 3.5 | 104.2 | 0.408 | 1.0, 5.7, 9.5 | 6.0 |
| | 43 | 200128 (9.41) | 1.8, 32.3, 5.0, 5.0 | 71.6 | 0.226 | 0.9, 5.7, 8.3, 10.0 | 6.0 |
| | 44 | 200329 (9.55) | 1.3, 22.3, 3.4, 1.5 | 47.0 | 0.210 | 3.1, 5.6, 8.2, 10.2 | 6.1 |
| | 45 | 200529 (9.70) | 1.1, 0.9, 20.8, 3.3, 1.8 | 48.4 | 0.259 | 0.7, 3.0, 5.6, 8.2, 10.4 | 6.2 |
| | 46 | 201018 (10.03) | 5.2, 4.1, 28.5, 7.3, 3.8 | 96.5 | 0.240 | −0.1, 3.5, 5.2, 8.7, 10.4 | 5.6 |
| | 47 | 201228 (10.20) | 6.6, 7.7, 30.5, 9.7 | 106.2 | 0.199 | −0.9, 3.5, 5.2, 8.7 | 5.1 |
| | 48 | 210222 (10.33) | 2.1, 5.7, 23.7, 1.5 | 74.2 | 0.166 | −1.3, 3.0, 5.2, 10.8 | 5.3 |
| | 49 | 210428 (10.49) | 1.5, 4.6, 14.0, 2.8, 1.0 | 39.2 | 0.332 | −1.4, 2.2, 5.2, 7.8, 9.5 | 4.7 |

Table 3
(Continued)

| Maser Line | Obs. Epoch | Date (Phase) (yymmdd) | $S_{peak}$ (Jy) | $\int S_\nu\, dv$ (Jy km s$^{-1}$) | rms (Jy) | $V_{peak}$ (km s$^{-1}$) | $V_{mean}$ (km s$^{-1}$) |
|---|---|---|---|---|---|---|---|
| | 50 | 210615 (10.60) | 3.8, 12.1, 2.0, 1.7 | 27.5 | 0.387 | 2.2, 5.2, 7.4, 8.2 | 5.0 |
| SiO129 | 11 | 120526 (2.76) | 4.7 | 5.1 | 0.520 | 6.6 | 6.4 |
| | 12 | 121211 (3.24) | … | … | … | … | … |
| | 13 | 130503 (3.58) | 6.0 | 7.3 | 0.433 | 6.6 | 6.7 |
| | 14 | 130917 (3.90) | 11.2 | 18.3 | 0.424 | 7.0 | 7.0 |
| | 15 | 131130 (4.08) | 3.4, 4.6, 8.2 | 36.9 | 0.485 | 5.4, 7.0, 8.0 | 7.1 |
| | 16 | 140214 (4.26) | … | … | … | … | … |
| | 17 | 140406 (4.38) | 3.6, 8.5 | 26.6 | 0.188 | 4.8, 7.0 | 7.0 |
| | 18 | 140604 (4.52) | … | … | … | … | … |
| | 19 | 150607 (5.40) | … | … | 0.416 | … | … |
| | 20 | 150818 (5.57) | … | … | … | … | … |
| | 21 | 151005 (5.68) | … | … | 0.263 | … | … |
| | 22 | 151213 (5.85) | 1.3, 1.0 | 3.1 | 0.163 | 6.2, 7.5 | 6.3 |
| | 23 | 160215 (6.00) | 1.2, 1.2 | 3.6 | 0.158 | 5.7, 7.0 | 6.5 |
| | 24 | 160405 (6.12) | 3.6, 5.0 | 12.7 | 0.152 | 5.7, 7.0 | 6.4 |
| | 25 | 160524 (6.23) | 1.9, 3.9 | 7.2 | 0.387 | 5.3, 7.0 | 6.8 |
| | 26 | 160920 (6.52) | 1.4, 2.0, 1.6 | 6.8 | 0.331 | 4.4, 6.1, 7.0 | 6.6 |
| | 27 | 161121 (6.66) | 0.8 | 3.6 | 0.233 | 4.5 | 5.8 |
| | 28 | 170131 (6.83) | 2.0, 1.9 | 5.3 | 0.204 | 5.3, 7.0 | 6.1 |
| | 29 | 170415 (7.01) | 7.4, 8.0 | 23.3 | 0.472 | 5.3, 6.6 | 5.9 |
| | 30 | 170603 (7.12) | 8.0, 1.4 | 22.1 | 0.388 | 6.2, 11.8 | 6.3 |
| | 31 | 170828 (7.32) | 2.6 | 1.7 | 0.632 | 6.2 | 6.2 |
| | 32 | 171101 (7.48) | 4.1 | 9.2 | 0.424 | 5.7 | 6.2 |
| | 33 | 180102 (7.62) | … | … | 0.387 | … | … |
| | 34 | 180308 (7.78) | 1.7, 2.0, 1.3, 0.8 | 3.95 | 0.302 | 4.8, 5.7, 7.0, 8.3 | 5.8 |
| | 35 | 180516 (7.94) | 17.7, 12.8, 9.4 | 55.2 | 0.687 | 5.3, 6.6, 9.2 | 6.7 |
| | 36 | 180909 (8.21) | 22.6, 14.5, 4.8 | 56.8 | 0.340 | 5.3, 6.6, 8.8 | 6.3 |
| | 37 | 181103 (8.34) | 18.2, 3.4, 1.0 | 37.9 | 0.302 | 5.3, 8.7, 10.5 | 6.2 |
| | 38 | 190106 (8.49) | 5.6, 2.0 | 12.4 | 0.263 | 5.7, 8.8 | 6.1 |
| | 39 | 190302 (8.62) | 0.9, 2.4, 1.2, 0.9 | 6.8 | 0.343 | 4.4, 5.7, 8.3, 9.6 | 6.3 |
| | 40 | 190514 (8.79) | 2.0, 3.8, 1.1, 1.0 | 7.6 | 0.529 | 4.9, 5.7 | 5.8 |
| | 41 | 190917 (9.09) | 14.7, 4.4, 2.8 | 39.5 | 0.597 | 5.7, 7.9, 9.7 | 6.3 |
| | 42 | 191113 (9.23) | 4.9, 1.3 | 9.4 | 0.551 | 5.7, 7.0 | 5.5 |
| | 43 | 200128 (9.41) | 8.7, 2.0, 1.1 | 17.1 | 0.328 | 5.7, 8.8, 10.0 | 6.3 |
| | 44 | 200329 (9.55) | 2.2, 3.8, 1.1, 1.0 | 9.7 | 0.328 | 4.9, 5.7, 7.9, 8.8 | 6.1 |
| | 45 | 200529 (9.70) | 3.2, 1.2 | 7.3 | 0.438 | 5.7, 8.3 | 5.8 |
| | 46 | 201018 (10.03) | 8.4, 1.2, 1.3 | 13.7 | 0.403 | 5.7, 7.9, 9.2 | 5.9 |
| | 47 | 201228 (10.20) | 7.4 | 13.9 | 0.232 | 5.7 | 5.7 |
| | 48 | 210222 (10.33) | 6.2 | 9.0 | 0.184 | 5.7 | 5.9 |
| | 49 | 210428 (10.49) | 4.7 | 4.5 | 0.519 | 5.7 | 5.7 |
| | 50 | 210615 (10.60) | … | … | 0.530 | … | … |
| $^{29}$SiO $v=0$ $J=1-0$ | 3 | 100227 (0.81) | … | … | 0.348 | … | … |
| | 4 | 100323 (0.87) | 1.6 | 3.1 | 0.258 | 6.6 | 6.7 |
| | 5 | 100504 (0.97) | 1.2 | 2.7 | 0.285 | 6.6 | 6.6 |
| | 6 | 100622 (1.08) | 1.5 | 1.7 | 0.274 | 6.5 | 6.6 |
| | 7 | 101012 (1.35) | … | … | 0.307 | … | … |
| | 8 | 111012 (2.22) | … | … | 0.287 | … | … |
| | 9 | 120103 (2.44) | … | … | 0.243 | … | … |

(This table is available in machine-readable form in the online article.)

It is clear that the OLC cycle and the overall variation pattern of the four SiO masers are correlated. However, the optical maximum phases of the central star and intensity maximum phases of the SiO masers exhibit slight phase lags and cycle-dependent variations. Table 5 shows the estimated phase lags for each cycle, along with the calculated mean and standard deviation (Std) values. The maximum points of each maser were estimated using the interpolation method shown in Figure 5.

The phase lags between the SiO masers and OLC exhibit the following range of values: 0.00–0.20 for SiO43, 0.03–0.18 for SiO42, 0.04–0.15 for SiO86, and 0.04–0.13 for SiO129. On average, the SiO43 and SiO42 masers exhibit a phase lag of 0.11 to the OLC, while the SiO86 and SiO129 masers show a slightly lower value of 0.07. It is widely recognized that a significant time delay (∼0.1–0.2) exists between changes in the optical brightness and SiO maser fluxes in the O-rich AGB stars (J. Alcolea et al. 1999; J. R. Pardo et al. 2004). The time delay can be attributed to the inherent spatial separation between the stellar photosphere, directly contributing to the star's visible surface, and the gas layers where SiO masers form. This delay can also be influenced by gradients in density and distance from the star. As we will discuss again in the next section, including the case of the H$_2$O maser, different regions of the CSE might respond differently to the shock, leading to a delay in the maser response compared to the optical pulsation changes. Additionally, there is a response time associated with maser mechanisms themselves. Even when the shock arrives at

Table 4
Observational Results of the H$_2$O Maser Line

| Epoch | Date (Phase) (yymmdd) | $S_{peak}$ B4, B3, B2, B1 (Jy) | $S_{peak}$ R1, R2 (Jy) | $\int S_\nu dv$ (Jy km s$^{-1}$) | rms (Jy) | $V_{peak}$ B4, B3, B2, B1 (km s$^{-1}$) | $V_{peak}$ R1, R2 (km s$^{-1}$) | $V_{mean}$ (km s$^{-1}$) |
|---|---|---|---|---|---|---|---|---|
| 1 | 090605 (0.18) | 36.5, ⋯, 16.4, 26.4 | ⋯, 16.4 | 138.5 | 0.457 | −1.9, ⋯, 1.9, 3.9, | ⋯, 8.6 | 2.5 |
| 2 | 091104 (0.54) | 1.4, ⋯, 1.3, ⋯ | 5.7, ⋯ | 13.9 | 0.274 | −2.5, ⋯, 1.5, ⋯ | 7.1, ⋯ | 5.9 |
| 3 | 100227 (0.81) | ⋯, ⋯, ⋯, ⋯ | 2.5, ⋯ | 4.5 | 0.296 | ⋯, ⋯, ⋯, ⋯ | 7.1, ⋯ | 7.7 |
| 4 | 100323 (0.87) | ⋯, 0.7, 1.6, ⋯ | ⋯, 3.0 | 8.7 | 0.252 | ⋯, −1.0, 1.5, ⋯ | ⋯, 8.3 | 6.9 |
| 5 | 100504 (0.97) | 5.9, ⋯, 7.0, ⋯ | ⋯, 4.5 | 31.5 | 0.267 | −1.8, ⋯, 1.9, ⋯ | ⋯, 8.7 | 2.2 |
| 6 | 100622 (1.08) | 44.0, ⋯, 28.0, ⋯ | ⋯, 10.0 | 152.0 | 0.340 | −1.8, ⋯, 1.5, ⋯ | ⋯, 8.3 | 1.2 |
| 7 | 101012 (1.35) | 6.8, ⋯, 18.7, ⋯ | 8.5, 7.2 | 83.5 | 0.347 | −2.1, ⋯, 1.7, ⋯ | 7.2, 8.5 | 2.8 |
| 8 | 111012 (2.22) | 20.5, ⋯, 42.8, ⋯ | 17.2, 17.8 | 176.5 | 0.261 | −2.5, ⋯, 1.7, ⋯ | 7.2, 8.5 | 2.5 |
| 9 | 120103 (2.44) | 2.6, 2.1, 12.8, 1.3 | 28.3, ⋯ | 75.9 | 0.166 | −2.4, −1.3, 1.7, 4.7 | 6.8, ⋯ | 4.6 |
| 10 | 120507 (2.72) | ⋯, ⋯, ⋯, ⋯ | 20.9, ⋯ | 26.0 | 0.253 | ⋯, ⋯, ⋯, ⋯ | 6.0, ⋯ | 6.2 |
| 11 | 120526 (2.76) | ⋯, ⋯, ⋯, ⋯ | 22.1, ⋯ | 25.8 | 0.291 | ⋯, ⋯, ⋯, ⋯ | 6.8, ⋯ | 7.0 |
| 12 | 121211 (3.24) | 24.3, ⋯, 19.0, ⋯ | 29.3, 16.8 | 159.4 | 0.134 | −2.5, ⋯, 1.3, ⋯ | 7.2, 8.9 | 2.5 |
| 13 | 130503 (3.58) | ⋯, ⋯, 2.2, ⋯ | 25.0, ⋯ | 34.8 | 0.254 | ⋯, ⋯, 1.8, ⋯ | 6.9, ⋯ | 6.3 |
| 14 | 130917 (3.90) | ⋯, 1.8, 3.8, 1.5 | 7.8, 10.8 | 22.1 | 0.298 | ⋯, −1.1, 1.8, 3.1 | 7.3, 9.0 | 4.2 |
| 15 | 131130 (4.08) | 16.2, ⋯, 31.1, ⋯ | 9.4, 10.0 | 125.5 | 0.219 | −1.5, ⋯, 1.1, ⋯ | 7.3, 9.0 | 1.7 |
| 16 | 140214 (4.26) | 17.3, ⋯, 28.9, ⋯ | 10.2, 10.1 | 130.0 | 0.173 | −1.5, ⋯, 1.1, ⋯ | 7.3, 9.0 | 1.6 |
| 17 | 140406 (4.38) | 4.8, ⋯, 4.4, ⋯ | 19.8, ⋯ | 96.9 | 0.204 | −1.5, ⋯, 1.1, ⋯ | 7.3, ⋯ | 2.3 |
| 18 | 140604 (4.52) | 19.1, ⋯, 15.7, ⋯ | 13.3, ⋯ | 49.6 | 0.347 | −1.5, ⋯, 1.0, ⋯ | 7.3, ⋯ | 4.4 |
| 19 | 150607 (5.40) | 16.7, ⋯, 13.8, ⋯ | 11.7, ⋯ | 74.9 | 0.291 | −2.4, ⋯, 1.1, ⋯ | 7.6, ⋯ | 1.5 |
| 20 | 150818 (5.57) | 2.8, ⋯, 2.6, ⋯ | 21.4, ⋯ | 33.6 | 0.290 | −2.4, ⋯, 2.2, ⋯ | 7.2, ⋯ | 5.0 |
| 21 | 151005 (5.68) | 1.3, 1.0, 1.2, ⋯ | 15.2, ⋯ | 20.8 | 0.202 | −2.4, −0.3, 1.8, ⋯ | 7.3, ⋯ | 5.9 |
| 22 | 151213 (5.85) | 1.9, ⋯, 2.4, ⋯ | 15.5, ⋯ | 24.3 | 0.183 | −2.4, ⋯, 1.8, ⋯ | 7.3, ⋯ | 5.2 |
| 23 | 160215 (6.00) | 22.9, ⋯, 16.3, ⋯ | 7.5, ⋯ | 79.7 | 0.173 | −2.4, ⋯, 1.5, ⋯ | 7.3, ⋯ | 1.0 |
| 24 | 160405 (6.12) | 35.6, ⋯, 21.4, ⋯ | 5.5, 7.7 | 111.3 | 0.181 | −2.4, ⋯, 1.5, ⋯ | 7.3, 8.6 | 0.7 |
| 25 | 160524 (6.23) | 42.5, ⋯, 26.1, ⋯ | 6.8, 9.9 | 134.0 | 0.249 | −2.4, ⋯, 1.5, ⋯ | 7.3, 8.6 | 0.7 |
| 26 | 160920 (6.52) | 3.7, ⋯, 2.8, ⋯ | 14.2, ⋯ | 23.0 | 0.227 | −2.4, ⋯, 1.4, ⋯ | 7.3, ⋯ | 4.9 |
| 27 | 161121 (6.66) | ⋯, ⋯, 0.8, ⋯ | 13.9, ⋯ | 14.9 | 0.193 | ⋯, ⋯, 1.4, ⋯ | 7.3, ⋯ | 7.0 |
| 28 | 170131 (6.83) | ⋯, ⋯, ⋯, ⋯ | 10.4, ⋯ | 10.9 | 0.179 | ⋯, ⋯, ⋯, ⋯ | 7.3, ⋯ | 7.4 |
| 29 | 170415 (7.01) | 10.9, ⋯, 11.8, ⋯ | ⋯, 10.8 | 56.6 | 0.282 | −2.8, ⋯, 1.0, ⋯ | ⋯, 8.6 | 2.4 |
| 30 | 170603 (7.12) | 24.3, 4.5, 23.5, ⋯ | ⋯, 18.7 | 96.2 | 0.236 | −2.8, −0.7, 1.0, ⋯ | ⋯, 8.6 | 1.9 |
| 31 | 170828 (7.32) | 13.9, 3.8, 21.0, 1.8 | ⋯, 7.8 | 65.5 | 0.305 | −2.8, −1.1, 1.0, 3.1 | ⋯, 8.6 | 1.6 |
| 32 | 171101 (7.48) | 3.1, 1.1, 10.9, 0.8 | 9.0, ⋯ | 31.0 | 0.238 | −2.6, −0.9, 1.2, 3.3 | 7.5, ⋯ | 3.1 |
| 33 | 180102 (7.62) | ⋯, ⋯, 3.4, ⋯ | 6.5, ⋯ | 8.9 | 0.245 | ⋯, ⋯, 1.0, ⋯ | 7.3, ⋯ | 5.8 |
| 34 | 180308 (7.78) | ⋯, ⋯, 2.6, ⋯ | 5.7, ⋯ | 10.1 | 0.178 | ⋯, ⋯, 1.2, ⋯ | 7.5, ⋯ | 5.4 |
| 35 | 180516 (7.94) | 2.0, 1.5, 18.4, ⋯ | 3.2, ⋯ | 32.8 | 0.287 | −2.4, −1.1, 1.0, ⋯ | 7.7, ⋯ | 1.9 |
| 36 | 180909 (8.21) | 17.5, ⋯, 59.4, ⋯ | ⋯, 10.9 | 138.2 | 0.244 | −3.0, ⋯, 0.8, ⋯ | ⋯, 8.8 | 1.5 |
| 37 | 181103 (8.34) | 9.1, ⋯, 50.8, ⋯ | 6.8, ⋯ | 91.0 | 0.171 | −3.0, ⋯, 0.8, ⋯ | 7.9, ⋯ | 1.5 |
| 38 | 190106 (8.49) | 1.5, ⋯, 21.8, ⋯ | 3.6, ⋯ | 33.9 | 0.142 | −3.0, ⋯, 0.8, ⋯ | 7.9, ⋯ | 2.2 |
| 39 | 190302 (8.62) | ⋯, ⋯, 9.0, 0.7 | 2.1, ⋯ | 11.9 | 0.171 | ⋯, ⋯, 0.8, 2.5 | 7.9, ⋯ | 2.7 |
| 40 | 190514 (8.79) | ⋯, ⋯, 8.4, ⋯ | 1.9, ⋯ | 10.2 | 0.221 | ⋯, ⋯, 0.8, ⋯ | 7.6, ⋯ | 2.6 |
| 41 | 190917 (9.09) | 41.5, ⋯, 120.8, ⋯ | 5.6, 7.2 | 224.6 | 0.288 | −3.0, ⋯, 0.8, ⋯ | 7.5, 8.4 | 0.3 |
| 42 | 191113 (9.23) | 31.6, ⋯, 105.4, ⋯ | 5.7, 6.5 | 202.9 | 0.283 | −3.0, ⋯, 0.8, ⋯ | 7.6, 8.4 | 0.4 |
| 43 | 200128 (9.41) | 4.7, ⋯, 46.1, ⋯ | ⋯, 6.0 | 70.6 | 0.177 | −3.0, ⋯, 0.7, ⋯ | ⋯, 8.4 | 1.3 |
| 44 | 200329 (9.55) | 0.8, ⋯, 10.5, ⋯ | 3.3, ⋯ | 17.6 | 0.190 | −3.0, ⋯, 0.8, ⋯ | 7.5, ⋯ | 2.8 |
| 45 | 200529 (9.70) | ⋯, ⋯, 1.9, ⋯ | 3.0, ⋯ | 5.6 | 0.246 | ⋯, ⋯, 0.4, ⋯ | 7.5, ⋯ | 5.7 |
| 46 | 201018 (10.03) | ⋯, ⋯, 16.0, ⋯ | 6.0, 6.0 | 32.5 | 0.200 | ⋯, ⋯, 0.8, ⋯ | 7.6, 8.4 | 4.0 |
| 47 | 201228 (10.20) | ⋯, ⋯, 19.5, ⋯ | 6.1, ⋯ | 35.3 | 0.165 | ⋯, ⋯, 0.4, ⋯ | 7.9, ⋯ | 3.3 |
| 48 | 210222 (10.33) | ⋯, ⋯, 12.4, ⋯ | 3.9, ⋯ | 21.1 | 0.190 | ⋯, ⋯, 0.4, ⋯ | 7.9, ⋯ | 3.3 |
| 49 | 210428 (10.49) | ⋯, 1.0, 4.0, ⋯ | 3.0, ⋯ | 13.9 | 0.249 | ⋯, −0.5, 0.8, ⋯ | 7.9, ⋯ | 4.8 |
| 50 | 210615 (10.60) | ⋯, ⋯, 0.9, ⋯ | 1.1, ⋯ | 4.3 | 0.305 | ⋯, ⋯, 0.8, ⋯ | 7.5, ⋯ | 7.1 |

(This table is available in machine-readable form in the online article.)

the maser-forming region, it takes some time for the masers to respond and emit radiation. The combination of these factors likely contributes to the observed time delay between changes in optical brightness and SiO maser fluxes.

In the context of SiO maser formation, the involvement of both collisional and radiative pumping mechanisms has been suggested (A. J. Kemball 2007, and references therein). The shock generated near the stellar surface, which emits optical radiation, propagates outward from the photosphere, influencing the inner layers of the CSE where the SiO molecules are located (under optimal density and temperature conditions for maser generation). At this point, maser emission will be generated as SiO molecules are collisionally pumped. The time delay of the SiO maser curve to the OLC can be considered as the time it takes to reach the SiO maser region after the occurrence of the shock (M. D. Gray et al. 2009, and references therein). On the other hand, if radiative pumping is dominant, the phase lag between the OLC and the maser curve would be very small or almost in the same phase. The strong correlation between the SiO flux density and IR radiation also supports this radiative pumping scenario (V. Bujarrabal et al. 1987; V. Bujarrabal 1994; J.-I. Nakashima & S. Deguchi 2007). As presented in Table 5, the phase lag of the SiO maser in this star appears to show a cycle-dependent variation. It is commonly understood that the generation of SiO masers is related to a combination of collisional and radiative pumping. However, it is not yet clear which of these mechanisms is dominant, and this may depend on the phase.

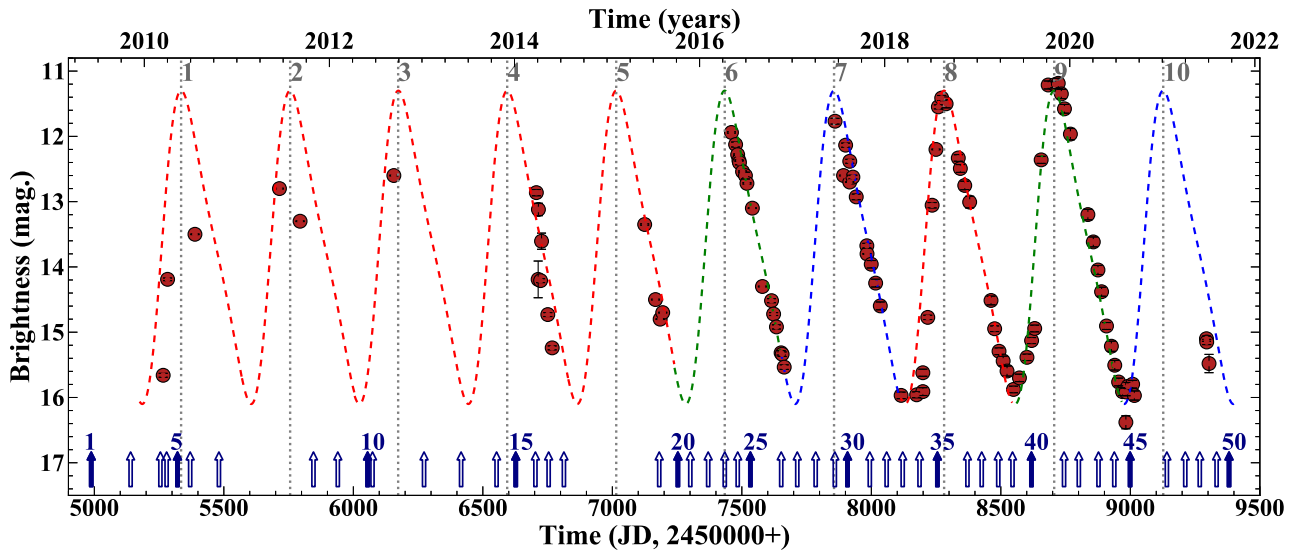

**Figure 3.** OLC of WX Ser from AAVSO (red circles), indicating a period of 421 ± 4 days based on the best fit. The observation epochs of the maser lines are marked with the blue arrows, and the relevant epoch numbers are presented. The optical maximum phases, marked with ordinal numbers, are represented by vertical dotted lines. Dashed lines represent the best-fitting tilted sinusoidal curves for individual cycles, distinguished by different colors (see text). Due to the lack of sufficient data points in the first five cycles, the same coefficients were used to determine the best-fit solution.

We compared our findings with the observational results of I. Gonidakis et al. (2013), who monitored the SiO43 maser during approximately three optical cycles of the Mira variable TX Cam using the Very Long Baseline Array (VLBA). The authors proposed that the shock waves generated by stellar pulsations during the second and third optical cycles of the TX Cam monitoring period directly influenced the maser intensity changes. In contrast, the shock waves generated during the first optical cycle did not increase the intensity of the maser, but rather exerted a dynamic effect of pushing the masing material outward. This led the authors to suggest that radiative pumping played a slightly more significant role in generating the SiO maser during the first optical cycle (when the phase lag between the SiO maser and the optical cycle was close to zero) than collisional pumping. Based on these observational results and analyses, we suggest that the phase lag range of WX Ser may be somewhat reduced when radiative pumping is relatively dominant. In addition, the maximum phases of SiO masers exhibit slight variations, gradually shifting from cycle to cycle. This phenomenon can be explained by the nonisotropic turbulence of gas and dust, as well as the convection of matter within the dust shell. These processes result in the propagation of multiple shock waves within the dust shell, leading to variations in the intensity and phase shift amplitudes (L. A. Nyman et al. 1986; B. Freytag et al. 2017). Therefore, the prevalence of the two mechanisms governing SiO maser generation changes with the surrounding conditions of the CSE.

Furthermore, we found a difference of ∼0.04 in the mean phase lag values between the SiO43/SiO42 pair and the SiO86/SiO129 pair, as shown in Tables 3 and 5. Considering this difference, J. F. Desmurs et al. (2014) reported that the ring structures of SiO masers with the same rotational level but different vibrational modes are formed in a similar region (roughly the same phase lag between the SiO43 maser and the SiO42 maser in our observations) around a central star. This result was derived through mapping observations of SiO $v=1$, 2, 3, and $J=1-0$ masers using the VLBA. In the case of different rotational level, R. Soria-Ruiz et al. (2004, 2007a) noted that the SiO43, SiO42, and SiO86 maser lines exhibit different ring distributions owing to the higher rotational level of the SiO86 maser, based on the VLBA images for $\chi$ Cyg, WX Psc, and R Leo. D.-H. Yoon et al. (2018) also reported minor variations in the ring distributions of all masers, through simultaneous VLBI imaging observation of SiO43, SiO42, SiO86, and SiO129 masers using the KVN for the supergiant VX Sgr. These results indicate that the variation in the average phase lags between the SiO43/SiO42 pair and the SiO86/SiO129 pair can be attributed to their different spatial distribution. However, our monitoring results show no difference in the phase delay between SiO86 and SiO129 masers (see Table 5). This may be explained by the sensitivity limits of our 129 GHz SiO maser observations; hence, SiO129 data are insufficient for a thorough investigation.

As shown in Figures 2(a)–(d) and 4(b)–(e), it appears that most of the detected SiO masers, especially the SiO43 and SiO42 masers, are somewhat scattered. According to the observation epochs, the SiO masers appear to be forming single peaks or spreading multiple peaks, indicating that the central star is highly active and pulsating. In addition, it seems that the peaks with the highest intensities are commonly located within approximately 2–3 km s$^{-1}$ of the $V_*$, which may be because the highest-velocity components are located farther from the central star and are more difficult to excite. This behavior is common to all four SiO masers. Furthermore, this phenomenon can also be seen in the velocity pattern of the masers. A more detailed discussion of the velocities will be given in Section 4.3.

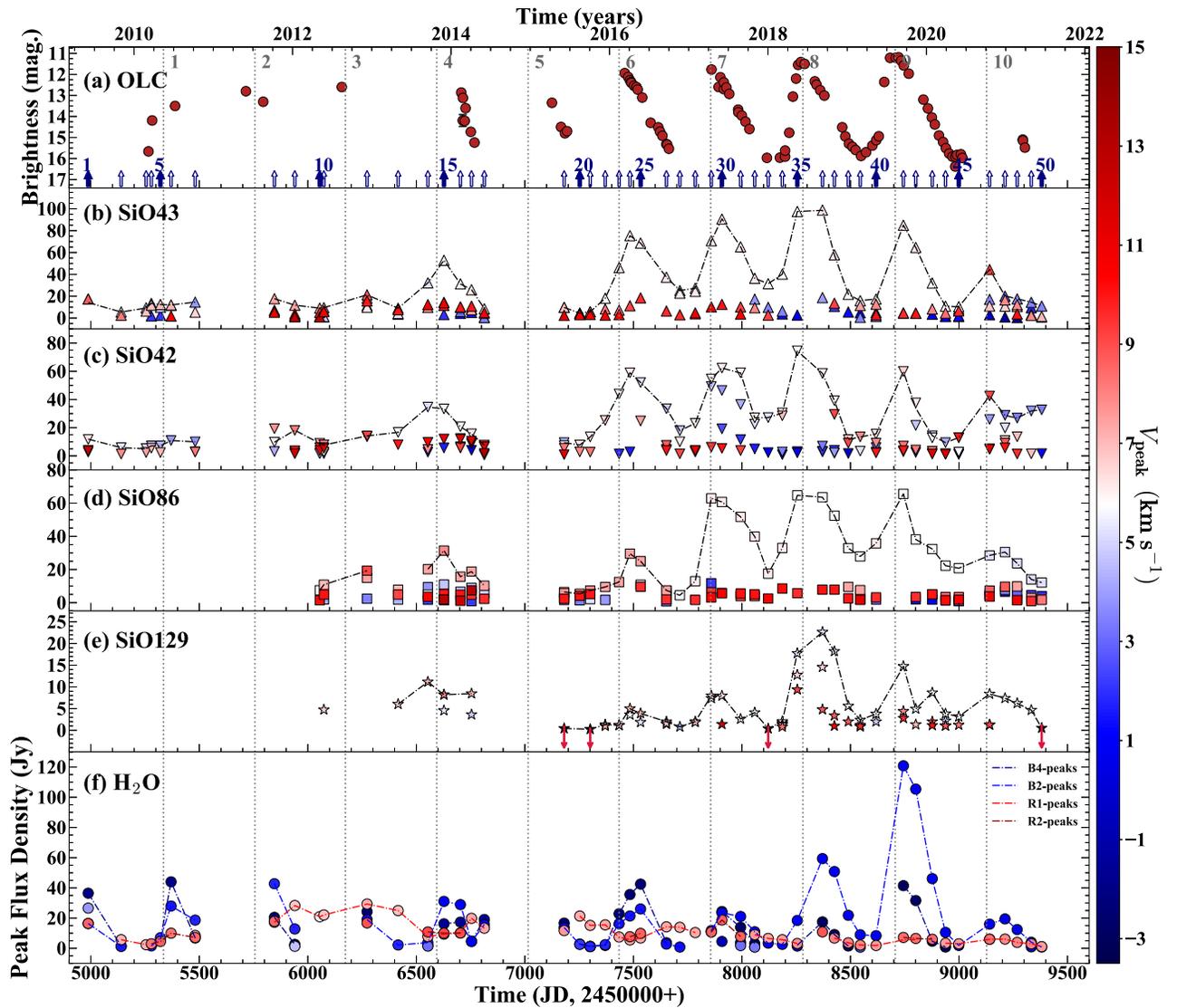

**Figure 4.** Peak flux densities of (b) SiO43, (c) SiO42, (d) SiO86, (e) SiO129, and (f) $H_2O$ masers of WX Ser, compared with (a) the OLC from 2009 June to 2021 June. The vertical dotted lines indicate the positions of the optical maxima. The blue arrows in panel (a) mark the monitoring epochs. For convenience, the optical cycles are numbered in gray, from the start of our monitoring. The downward-pointing red arrows represent the upper limit ($3\sigma$) of the epochs that were not detected. The highest peaks of SiO masers are presented as dotted–dashed black lines in panels (b)–(e). The color bar on the right provides an indication of the true peak velocities of masers.

#### 4.2.2. Characteristics of the $H_2O$ Maser Line

The plot illustrated in Figures 4(f) and 5(f) shows that, as with the SiO masers, the intensity variation of the $H_2O$ maser correlates very well with the OLC pattern. With the exception of a massive flare event in the ninth cycle, the pattern of intensity variation over time for the $H_2O$ maser is remarkably similar to that of the SiO masers, that is, the intensity of the $H_2O$ maser remains gently below the peak flux density of $\sim$50 Jy, being also correlated well with the OLC, and then increases drastically from the eighth optical maximum (at epoch 36), exhibiting a strong flare near the ninth optical maximum (at epoch 41), as described in Section 3.

As shown in Figures 1 and 4(f), the $H_2O$ maser of WX Ser exhibits mainly four distinct peak components, the blueshifted B4 and B2 peaks and the redshifted R1 and R2 peaks.

Furthermore, the blueshifted components display significantly higher flux densities than the redshifted components. They also present a more asymmetric distribution in relation to the $V_*$. This asymmetry was consistently observed during our monitoring period. Notably, with the exception of epoch 12 ($\phi = 3.24$), the B-peaks exhibited the highest intensity levels in most maximum phases, even during the flaring event in epoch 41 ($\phi = 9.09$). The dominance of the blueshifted peaks and their higher intensity levels suggest that the blueshifted direction potentially represents more energetic or prolific maser-producing regions in the surrounding environment of WX Ser. Additionally, the consistent asymmetry in maser distribution implies that kinematic motion or structural asymmetries in the maser-emitting regions favor the blueshifted direction of this star.

The phase lags between the $H_2O$ maser and OLC phase are greater than those of the SiO masers, as shown in Table 5 and Figure 5. The average phase lag of the $H_2O$ maser (relative to the OLC phase) is $\sim$0.17, with a range of 0.12–0.23, and approximately 0.06–0.1 larger than the phase lags for the SiO masers. The difference in phase lags between the SiO and $H_2O$ masers may be due to their inherent differences in terms of the formation region and pumping mechanism. It is known that $H_2O$ masers are mainly found in outer layers where dust is forming, typically associated with the outer regions of CSEs. In contrast, SiO masers are predominantly located within the inner regions of CSEs in relation to dust layers. This distinction in maser distribution points to varied physical conditions in densities, kinetic temperatures, and dynamic motions (H. J. Habing 1996; H. Imai et al. 2003), as well as processes occurring in these different regions of the circumstellar environment. The occurrence of the $H_2O$ maser is attributable to the collisional excitation and de-excitation of $H_2O$ molecules, unlike the the SiO masers, which are influenced by a combination of radiative and collisional pumping (A. J. Kemball 2007).

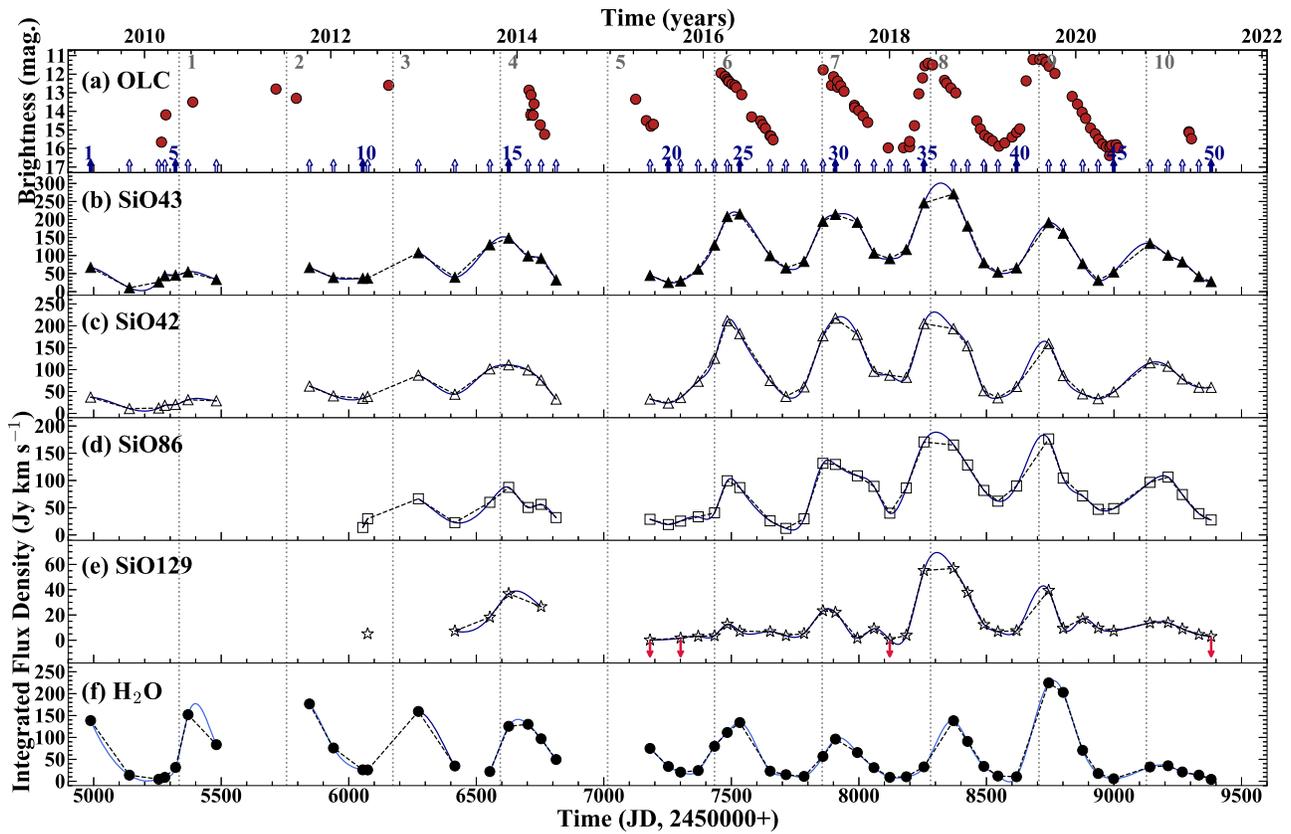

**Figure 5.** Integrated flux densities of SiO43, SiO42, SiO86, SiO129, and $H_2O$ masers of WX Ser, compared with the OLC from 2009 June to 2021 June. Blue solid lines represent the curves fitted with a B-spline for the observed data. The vertical dotted lines and blue and red arrows are the same as depicted in Figure 4.

Meanwhile, as briefly discussed above, during most of the observed period, the peak flux density of the $H_2O$ maser has generally remained consistent below ∼50 Jy and in good agreement with the variation in the OLC. However, it exhibits a sudden increased emission in epoch 41 ($\phi = 9.09$) owing to the flaring component, as depicted in Figures 2(e)–(f) and 4. At this epoch, the B4, B2, R1, and R2 components reach peak flux densities of ∼41, ∼120, ∼5, and ∼7 Jy, respectively. The B2 component, detected within a limited velocity range ($V_{LSR} = 0\text{–}2$ km s$^{-1}$), indicates runaway phenomena characterized by a sudden surge in flux density, reaching values 5–6 times higher than the average. Subsequently, it rapidly diminishes while maintaining a consistent line width. This observation suggests that the $H_2O$ cloud, located in a localized and limited area, satisfies conditions for short-term amplitude augmentation of the maser (A. M. S. Richards et al. 2011). Therefore, when considering the overall temporal variation trend of the $H_2O$ maser in WX Ser, it is speculated that the $H_2O$ maser clumps are sparsely distributed throughout the CSE, forming a "density-bounded cloud," a concept introduced by J. Brand et al. (2020). The material at the inner boundary of the $H_2O$ maser region, located just outside the dust layer or inner side of its masing region, may experience density enhancements due to multiple periodic pulsation-driven shock waves. Following the passage of a strong shock wave through these bounded clouds, the intensity of the maser likely increases compared to its previous state. These shock waves are generated by the periodic pulsation of the central star, and they create density perturbations that can cause the $H_2O$ maser to become more active. This is because the density fluctuations can produce additional collisions between molecules, which can lead to an increased production of radiation.

Notably, simultaneous observations of SiO and $H_2O$ masers are crucial for tracing the locations of the hypothesized shock waves. As shown in Figures 4(f) and 5(f), the flux density of the $H_2O$ maser, in conjunction with the SiO masers, shows a moderate fluctuation without sudden, drastic changes. It then appears to undergo a significant exponential increase in intensity during the eighth and ninth cycles, indicating the massive flare event. In this case, the SiO masers exhibit the highest flux density in the eighth cycle, whereas the $H_2O$ maser demonstrates the highest flux density in the ninth cycle, following a temporal pattern. This suggests a potential temporal relationship between the material flow during the flare event and the behavior of these two types of masers. It has been

Table 5
Maximum Phases and Phase Lags of Each Maser Line Relative to the OLC

| OLC Time | $H_2O$ | | SiO43 | | SiO42 | | SiO86 | | SiO129 | |
|---|---|---|---|---|---|---|---|---|---|---|
| | Max. | Lag | Max. | Lag | Max. | Lag | Max. | Lag | Max. | Lag |
| 5335 | 5398 | 0.15 | 5383 | 0.11 | 5403 | 0.16 | ⋯ | ⋯ | ⋯ | ⋯ |
| 5756 | ⋯ | ⋯ | ⋯ | ⋯ | ⋯ | ⋯ | ⋯ | ⋯ | ⋯ | ⋯ |
| 6173 | 6269 | 0.23 | 6256 | 0.20 | 6249 | 0.18 | 6203 | 0.07 | ⋯ | ⋯ |
| 6594 | 6663 | 0.17 | 6607 | 0.03 | 6623 | 0.07 | 6618 | 0.06 | 6615 | 0.05 |
| 7015 | ⋯ | ⋯ | ⋯ | ⋯ | ⋯ | ⋯ | ⋯ | ⋯ | ⋯ | ⋯ |
| 7434 | 7529 | 0.23 | 7511 | 0.19 | 7494 | 0.15 | 7489 | 0.13 | 7489 | 0.13 |
| 7856 | 7921 | 0.16 | 7939 | 0.20 | 7926 | 0.17 | 7877 | 0.05 | 7882 | 0.06 |
| 8281 | 8366 | 0.19 | 8322 | 0.09 | 8296 | 0.03 | 8300 | 0.04 | 8305 | 0.05 |
| 8706 | 8759 | 0.12 | 8750 | 0.10 | 8728 | 0.05 | 8724 | 0.04 | 8724 | 0.04 |
| 9127 | 9187 | 0.14 | 9129 | 0.00 | 9164 | 0.09 | 9191 | 0.15 | 9182 | 0.13 |
| Mean | | 0.17 | | 0.11 | | 0.11 | | 0.07 | | 0.07 |
| Std | | 0.04 | | 0.07 | | 0.05 | | 0.04 | | 0.04 |

**Note.** "Time" indicates JD+2450000.

frequently noted that $H_2O$ masers rarely produce sudden flares. Fortunately, we successfully monitored a singular $H_2O$ flare event both before and during its occurrence. This enabled us to deduce whether there was any temporal trend in the intensity variations of the SiO and $H_2O$ masers leading up to and during the flare, as follows. It is expected that the shocks induced by the periodic stellar pulsations may have a long-lasting impact on the SiO masing region, leading to an increase in the SiO maser intensity. Despite the lack of data for the fifth OLC in our monitoring series, it seems plausible to assume that this increasing trend would continue until the eighth or ninth OLC. Based on our observations, it appears that after reaching the SiO masing region the shocks may propagate to the $H_2O$ masing region with a relative time lag of about one cycle. This could explain the sudden increase in the intensity of the $H_2O$ maser from the eighth cycle stage and its peak in the ninth cycle (by this time, the SiO maser has already reached its maximum intensity). In other words, there seems to be a time lag of approximately one cycle between the SiO and $H_2O$ masing regions when a significant flow of material is involved in the production of a flare.

However, this hypothesis has several constraints that make it far from clear. Since there is a dust-forming layer between the SiO and $H_2O$ masing regions, its effects must be taken into account. Moreover, as the SiO maser is bounded by the dust-forming layer and displays turbulent dynamics with predominantly random outward and inward motion, it is not yet possible to definitively conclude that there is a time delay of one cycle. Furthermore, it would be a mistake to generalize the results of a single $H_2O$ maser flare event to discuss the time delay between SiO and $H_2O$ masers caused by a shock wave. In this case, we simply consider the possibility of a one-cycle time delay without taking all constraints into account.

Regarding the spatial separation between the two maser regions with respect to the stellar surface, it is highly probable that the B components mainly originate from the inner shell radius of the $H_2O$ maser region ($\sim 5R_*$). This notion is supported by the significant variability detected in the blueshifted line (G. M. Rudnitskii & A. A. Chuprikov 1990; A. M. S. Richards et al. 2011; H. Imai et al. 2019). In contrast, the excitation of the SiO maser occurs within $\sim 2R_*$–$4R_*$ (P. J. Diamond et al. 1994). The $R_*$ of WX Ser has been estimated to be $3.6 \times 10^{13}$ cm, based on the dust shell model proposed by K. Justtanont & A. G. G. M. Tielens (1992). This implies that the distance between the SiO and $H_2O$ maser regions is 2.4–7.2 au. Using the velocity at the blue edge of the $H_2O$ maser spectra (see Section 4.3.2.), the shock velocity is estimated to be approximately 10.5 km s$^{-1}$, consistent with the theoretical framework proposed by G. M. Rudnitskii & A. A. Chuprikov (1990). As a result, the shock travel time in WX Ser is approximately one to two cycles. In comparable instances, the SiO86 maser intensity in U Ori showed a significant increase in 1982 December, as reported by L. A. Nyman & H. Olofsson (1986). This event is linked to the observation of the $H_2O$ maser reported by G. M. Rudnitskij et al. (2000), who detected a 2.5 times higher intensity of the $H_2O$ maser in U Ori during 1983 December compared with the previous maximum phase. J. Alcolea et al. (1999) reported the gradual increase in the intensities of the SiO43 and SiO42 masers in W Hya between 1987 April and 1989 May. This observation is consistent with the results reported by G. M. Rudnitskii et al. (1999), who recorded an increase in the intensity of the $H_2O$ maser in W Hya in the period from 1988 July to 1991 October. Considering these observations, we propose that shock waves propagate from the SiO maser region to the $H_2O$ maser region in WX Ser over a period of roughly one to two cyclic periods.

From these analyses, the sequential variations in the enhanced flux density of the $H_2O$ maser can be attributed to the accumulation of $H_2O$ molecules in the CSE. Furthermore, we investigated the potential relationship between the SiO maser region and flaring of $H_2O$ masers, focusing on the influence of pulsation-driven shock waves. To strengthen confidence in the aforementioned hypotheses, it is crucial to carry out simultaneous VLBI observations of SiO and $H_2O$ masers directed toward WX Ser.

### 4.2.3. Maser Intensity Ratios

The intensity ratio of the masers can be another key factor that helps identify differences between various molecules, specifically $H_2O$ and SiO molecules, and between different rotational lines in a given $v$-state of the SiO line, namely SiO43, SiO86, and SiO129 (e.g., J. Alcolea et al. 1999; S.-H. Cho et al. 2014; J. Kim et al. 2014; M. C. Stroh et al. 2018; H. Yang et al. 2021; D.-H. Yoon et al. 2023). In particular, a direct comparison of the intensity ratio between $H_2O$ and SiO masers is only feasible with the KVN system, which enables the simultaneous observation of these two masers. This represents a significant advancement in the field, as it was previously not possible to conduct such a comparative study.

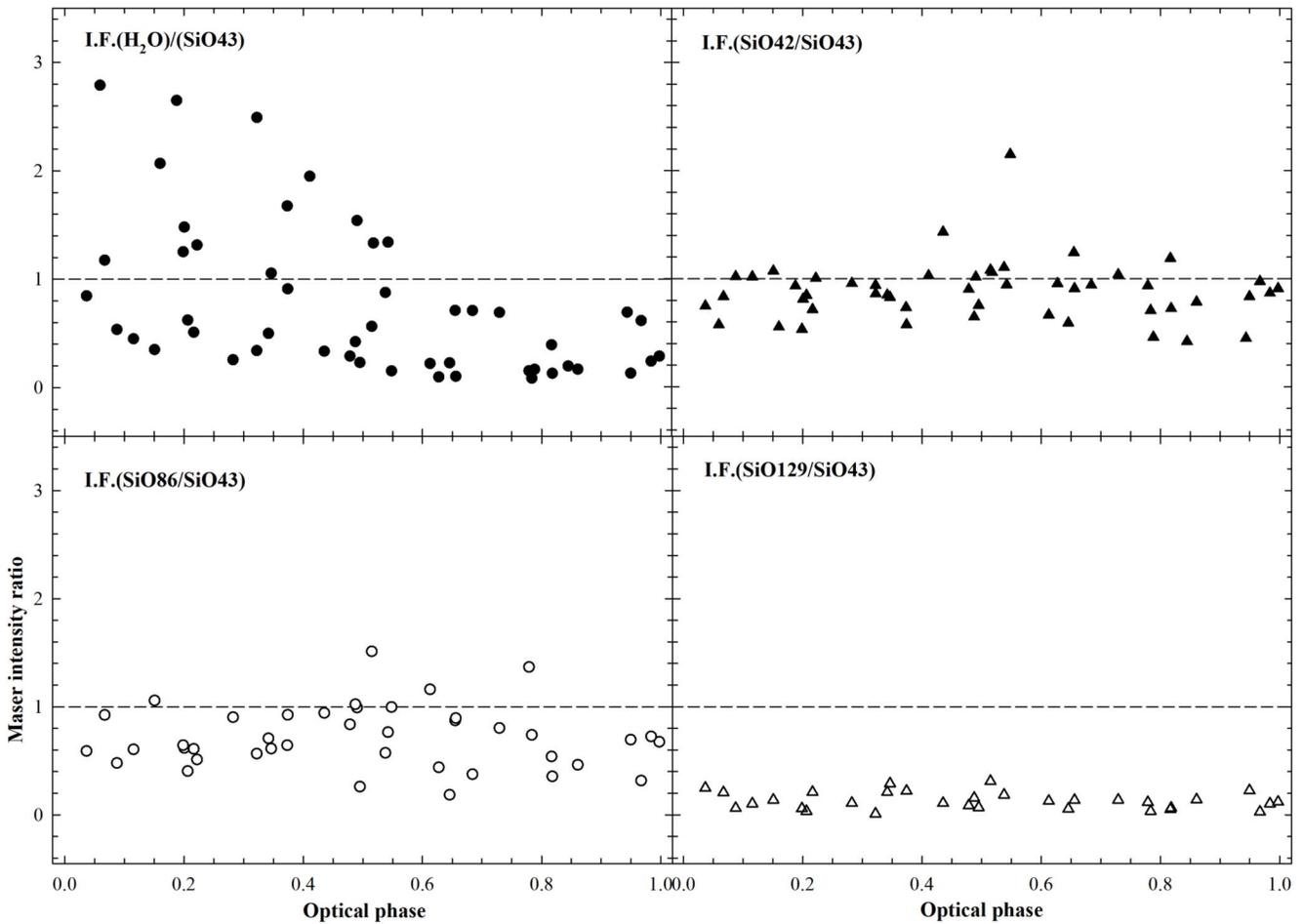

**Figure 6.** Integrated flux density (I.F.) ratios of the $H_2O$, SiO42, SiO86, and SiO129 masers to the SiO43 maser according to the normalized optical phase.

Figure 6 presents a comparison of the integrated flux density ratios (I.F.) of $H_2O$, SiO42, SiO86, and SiO129 masers to the SiO43 maser, namely, I.F.($H_2O$/SiO43), I.F.(SiO42/SiO43), I.F.(SiO86/SiO43), and I.F.(SiO129/SiO43), as outlined in Table 6. All optical phases have been normalized from 0.0 to 1.0. Table 6 also includes the intensity ratio of SiO129 to SiO86, I.F.(SiO129/SiO86), to compare with previous studies. The mean value and standard deviation for each ratio are presented at the bottom of Table 6. It is noteworthy that the line profiles and intensities of the detected masers exhibit considerable variability depending on the observation epoch, which in turn leads to relatively large standard deviations compared to the mean values of these ratios.

The I.F.(SiO42/SiO43) and I.F.(SiO86/SiO43) of WX Ser are in good agreement with previously obtained values from various surveys and monitoring. Meanwhile, for I.F.(SiO129/SiO43), there is currently no previous comparative study available, except for simultaneous observations with the KVN. An exemplary study on this ratio is the aforementioned one conducted by S.-H. Cho et al. (2014). Through the analysis of their observational results, the mean I.F.(SiO129/SiO43) ≈ 0.39 was determined for TX Cam, with SiO43 detected significantly stronger than SiO129 in all epochs used in the analysis (seven epochs). Furthermore, this ratio was also derived from 8 yr of monitoring observations on V627 Cas, a star classified as a Mira variable, albeit recently suspected to be a symbiotic star (H. Yang et al. 2021). The results of their analyses were reported in terms of antenna temperature units. To enhance clarity, we converted them all to units of flux density and recalculated the I.F. ratios. This resulted in a value of 0.12. H. Yang et al. (2021) also examined the intensity ratio of SiO129/SiO86 and found an average value of 0.13 after converting to flux density units. In addition, we calculated the ratio, I.F.(SiO129/SiO86), for R Cas, a representative Mira variable star, from the work by J. Kang et al. (2006), who conducted simultaneous monitoring observations of SiO86 and SiO129. The average value was determined to be 0.15 (after converting to flux density units). Both of these values are similar to our results.

The intensity ratios of SiO masers are dependent on the predominant pumping mechanism and vary depending on its energy level. In response to this complexity, M. D. Gray et al. (2009) developed a unified model of SiO maser pumping by addressing several weaknesses in previous models, including multiple layers in CSE, an accurate estimation of the IR radiation field from dust, sufficient chemical abundances for particles capable of colliding with SiO, and new rate coefficients for SiO + H collisions, as well as the relationship between the model phase and the OLC. We have compared the trend in intensity ratios along the SiO maser transitions in

Table 6
Maser Intensity Ratios

| Epoch | Date (yymmdd) | Optical Phase ($\phi$) | $\frac{\text{I.F.}(H_2O)}{\text{I.F.}(SiO43)}$ | $\frac{\text{I.F.}(SiO42)}{\text{I.F.}(SiO43)}$ | $\frac{\text{I.F.}(SiO86)}{\text{I.F.}(SiO43)}$ | $\frac{\text{I.F.}(SiO129)}{\text{I.F.}(SiO43)}$ | $\frac{\text{I.F.}(SiO129)}{\text{I.F.}(SiO86)}$ |
|---|---|---|---|---|---|---|---|
| 1 | 090605 | 0.18 | 2.07 | 0.55 | ... | ... | ... |
| 2 | 091104 | 0.54 | 1.33 | 1.06 | ... | ... | ... |
| 3 | 100227 | 0.81 | 0.17 | 0.46 | ... | ... | ... |
| 4 | 100323 | 0.87 | 0.20 | 0.42 | ... | ... | ... |
| 5 | 100504 | 0.97 | 0.70 | 0.45 | ... | ... | ... |
| 6 | 100622 | 0.08 | 2.79 | 0.57 | ... | ... | ... |
| 7 | 101012 | 0.35 | 2.49 | 0.86 | ... | ... | ... |
| 8 | 111012 | 0.22 | 2.65 | 0.93 | ... | ... | ... |
| 9 | 120103 | 0.44 | 1.95 | 1.03 | ... | ... | ... |
| 10 | 120507 | 0.72 | 0.71 | 0.94 | 0.38 | ... | ... |
| 11 | 120526 | 0.76 | 0.69 | 1.04 | 0.80 | 0.14 | 0.17 |
| 12 | 121211 | 0.24 | 1.48 | 0.81 | 0.62 | ... | ... |
| 13 | 130503 | 0.58 | 0.88 | 1.10 | 0.57 | 0.19 | 0.32 |
| 14 | 130917 | 0.90 | 0.17 | 0.78 | 0.46 | 0.14 | 0.31 |
| 15 | 131130 | 0.08 | 0.85 | 0.75 | 0.59 | 0.25 | 0.42 |
| 16 | 140214 | 0.26 | 1.32 | 1.00 | 0.51 | ... | ... |
| 17 | 140406 | 0.38 | 1.06 | 0.83 | 0.61 | 0.29 | 0.47 |
| 18 | 140604 | 0.52 | 1.54 | 1.02 | 0.99 | ... | ... |
| 19 | 150607 | 0.40 | 1.68 | 0.73 | 0.64 | ... | ... |
| 20 | 150818 | 0.57 | 1.34 | 0.94 | 0.76 | ... | ... |
| 21 | 151005 | 0.68 | 0.71 | 1.24 | 0.87 | ... | ... |
| 22 | 151213 | 0.85 | 0.39 | 1.19 | 0.54 | 0.05 | 0.09 |
| 23 | 160215 | 0.00 | 0.62 | 0.98 | 0.32 | 0.03 | 0.09 |
| 24 | 160405 | 0.12 | 0.54 | 1.02 | 0.48 | 0.06 | 0.13 |
| 25 | 160524 | 0.23 | 0.62 | 0.85 | 0.40 | 0.03 | 0.08 |
| 26 | 160920 | 0.52 | 0.23 | 0.75 | 0.26 | 0.07 | 0.26 |
| 27 | 161121 | 0.66 | 0.23 | 0.59 | 0.18 | 0.05 | 0.30 |
| 28 | 170131 | 0.83 | 0.13 | 0.72 | 0.35 | 0.06 | 0.18 |
| 29 | 170415 | 0.01 | 0.29 | 0.91 | 0.68 | 0.12 | 0.18 |
| 30 | 170603 | 0.12 | 0.45 | 1.02 | 0.61 | 0.10 | 0.17 |
| 31 | 170828 | 0.32 | 0.34 | 0.94 | 0.57 | 0.01 | 0.02 |
| 32 | 171101 | 0.48 | 0.29 | 0.90 | 0.84 | 0.09 | 0.10 |
| 33 | 180102 | 0.62 | 0.10 | 0.96 | 0.44 | ... | ... |
| 34 | 180308 | 0.78 | 0.09 | 0.71 | 0.74 | 0.03 | 0.05 |
| 35 | 180516 | 0.94 | 0.13 | 0.84 | 0.69 | 0.22 | 0.32 |
| 36 | 180909 | 0.21 | 0.51 | 0.71 | 0.61 | 0.21 | 0.34 |
| 37 | 181103 | 0.34 | 0.50 | 0.85 | 0.71 | 0.21 | 0.29 |
| 38 | 190106 | 0.49 | 0.42 | 0.65 | 1.02 | 0.16 | 0.15 |
| 39 | 190302 | 0.62 | 0.22 | 0.67 | 1.16 | 0.13 | 0.11 |
| 40 | 190514 | 0.79 | 0.16 | 0.93 | 1.37 | 0.12 | 0.08 |
| 41 | 190917 | 0.09 | 1.18 | 0.83 | 0.92 | 0.21 | 0.22 |
| 42 | 191113 | 0.23 | 1.25 | 0.53 | 0.64 | 0.06 | 0.09 |
| 43 | 200128 | 0.41 | 0.91 | 0.58 | 0.92 | 0.22 | 0.24 |
| 44 | 200329 | 0.55 | 0.57 | 1.08 | 1.51 | 0.31 | 0.21 |
| 45 | 200529 | 0.70 | 0.10 | 0.91 | 0.89 | 0.14 | 0.15 |
| 46 | 201018 | 0.03 | 0.24 | 0.87 | 0.72 | 0.10 | 0.14 |
| 47 | 201228 | 0.20 | 0.35 | 1.07 | 1.06 | 0.14 | 0.13 |
| 48 | 210222 | 0.33 | 0.26 | 0.96 | 0.90 | 0.11 | 0.12 |
| 49 | 210428 | 0.49 | 0.34 | 1.43 | 0.94 | 0.11 | 0.11 |
| 50 | 210615 | 0.60 | 0.15 | 2.15 | 1.00 | ... | ... |
| Mean | | | 0.77 | 0.88 | 0.71 | 0.13 | 0.19 |
| Std | | | 0.69 | 0.27 | 0.28 | 0.08 | 0.11 |

(This table is available in machine-readable form in the online article.)

WX Ser with the model of M. D. Gray et al. (2009) and found that the SiO masers in WX Ser fit best with model 4 (the hydrodynamic solution M21n), based on the pattern of peak amplification factors for each model shown in their Figure 10. In this model, stellar radiation and kinematic collisions are the only sources of pumping. As the pumping mechanism of these masers is also closely related to the ring structure of the SiO masers, we plan to conduct VLBI observations in future studies to obtain a spatial distribution map of the SiO masers.

In the case of I.F.($H_2O$/SiO43), no system was available to observe these two masers simultaneously prior to the development of the KVN. Therefore, we believe that the most reliable data come from our previous studies, which utilized this system to perform simultaneous observations of $H_2O$ and SiO masers since 2009, when the KVN became operational (J. Kim et al. 2010, 2013; S.-H. Cho & J. Kim 2012). J. Kim et al. (2014) analyzed the ratios of these two masers for 111 Mira variables, yielding a mean value of 0.70, which is highly similar to the results obtained in the present study.

The variation pattern of I.F.($H_2O$/SiO43) is easily distinguished from the ratio of the other SiO masers to the SiO43 maser. Namely, most of the observed data with an I.F.($H_2O$/SiO43) > 1 are distributed between $\phi = 0.0$ and 0.5. In contrast, the ratios of SiO42, SiO86, and SiO129 to SiO43 are lower than 1 in most optical phases, regardless of the phase. As previously mentioned, the varying intensity ratio of the $H_2O$ to the SiO43 maser, which is correlated with pulsation phase (namely the difference in sensitivity of the $H_2O$ maser to the optical phase compared to the SiO maser), can be attributed to the different excitation mechanisms and physical conditions of the regions where they occur.

The $H_2O$ maser is primarily formed in gas–dust regions in the inner layers of CSE. The presence of this maser is closely linked to the mass-loss processes occurring in the CSE, which can be significantly influenced by the overall stellar pulsation. During the optical maximum phase, the star undergoes a period of enhanced mass loss, leading to an increase in density and physical conditions favoring the production of the $H_2O$ maser. Therefore, the intensity of the $H_2O$ maser is stronger during the optical maximum phase, due to the favorable conditions present in the inner layers of the CSE.

On the other hand, SiO masers predominantly emanate from the inner atmosphere of Mira variables, closer to the stellar photosphere. Due to the nature of population inversion, the excitation conditions necessary for SiO masers to occur are similar across phases within the same vibrational state ($v = 1$). This phenomenon is observed to manifest similarly across various $J$-levels (R. Soria-Ruiz et al. 2007b). As a result of this consistency in excitation conditions, the intensity ratios of SiO42, SiO86, and SiO129 to SiO43 tend to remain relatively stable across the different phases. The stability of these intensity ratios results in the intensity ratios of SiO42, SiO86, and SiO129 relative to SiO43 appearing to be independent of the phase, as compared to the phase-sensitive response of I.F.($H_2O$/SiO43).

### 4.3. Time Variations of SiO and $H_2O$ Maser Velocities

The temporal variation of the velocities of the SiO and $H_2O$ masers of Mira variable stars in general highlights the intricate interplay between stellar pulsations, mass loss, and circumstellar environments during the thermally pulsing AGB (TP-AGB) phase. In the Mira variable star, a representative TP-AGB phase object, the SiO maser shows velocities corresponding to turbulence with phase-dependent infall and outflow motions within a range of approximately 10 km s$^{-1}$ relative to the $V_*$, while the $H_2O$ maser shows velocity extents of up to about 40 km s$^{-1}$ relative to the $V_*$, attributable to the influence of the accelerating stellar wind (J. Kim et al. 2014). Practically, we can infer what kinematic/dynamic motion is currently occurring in the surrounding atmosphere and in the CSE of this star from a detailed analysis of the time variations of these masers in WX Ser.

#### 4.3.1. Peak and Mean Velocities of SiO and $H_2O$ Masers

The velocity trends of the peak flux components of the SiO masers, $V_{peak}$, are shown in Figures 7(b)–(e), with the OLC in Figure 7(a). Included are all peak components shown in Table 3. The color bar on the right side of the figure shows the range of the color scale that changes continuously over time in accordance with each peak velocity. In the same order, Figures 8(b)–(e) show the time variations of the $V_{mean}$ with respect to $V_*$, and Figure 8(a) shows the OLC.

As illustrated in Figure 7, the velocity distributions of SiO masers do not exhibit any discernible systematic variation over time. Additionally, they appear to be slightly more dispersed than the velocity distributions of SiO masers in widely known Mira variable stars. Representative samples of these expanded line profiles can be seen in epochs 15–17 and 45–48 of Figure 1. It seems that this multipeaked line profile is a common feature in SiO maser spectra of WX Ser, regardless of the observed phase. However, the most prominent peak components in the majority of spectra are typically found within a velocity range of −6 to 4 km s$^{-1}$ relative to the $V_*$. The amplification of SiO masers is exponentially dependent on the effective and coherent path length along the line of sight. This makes the maser lines particularly sensitive to the prevailing velocity field. Hence, masers originating from these specific regions exhibit a notable tendency for tangential amplification close to the $V_*$ at the optical maximum phase.

In addition, the full width of the maser line profiles, or velocity width, may be subject to broadening based on the variability of the OLC. This broadening effect is thought to be directly linked to stellar pulsation, convective motion, and possible rotational activity, all of which contribute to the observed line profiles. These dynamics are effectively depicted in Figures 7 and 8, illustrating the time-varying velocity features of SiO maser emissions. The observed spectrum with the largest velocity width among the SiO masers is from epoch 17 of SiO43. The velocity width was measured to be up to ∼18 km s$^{-1}$, as we discuss again in the next section. If we consider the moderated value of this velocity width to be the terminal velocity, we can estimate that the terminal velocity of the SiO43 would be ∼9 km s$^{-1}$. It is worth noting that this value is slightly higher than the known expansion velocity of WX Ser, which has been estimated to be approximately 7.6 km s$^{-1}$ (J. Herman & H. J. Habing 1985). This velocity excess indicates that the kinematic velocity of the gas in the SiO maser region may permit velocity surpassing the terminal velocity, although the SiO maser generates in the innermost region of the shell, where it is expected to be similar to the $V_*$. As was previously indicated, in this case, a number of kinematic phenomena, such as turbulent motion, gas input and outflow, rotation, and asymmetric mass ejection, could contribute to the formation of SiO masers.

On the other hand, the velocity trends of the SiO86 and SiO129 masers with time do not show any noticeable differences compared to those of SiO43 and SiO42. The velocity distributions of SiO86 and SiO129 are similar to those of SiO43 and SiO42, with the main-peak component exhibiting the strongest intensity, primarily situated in the velocity region near the $V_*$. Subpeak components with weaker intensities intermittently appear in the red- and blueshifted regions relative to the $V_*$. However, in epochs 35–36, which exhibited the strongest emission for all SiO masers during the entire monitoring period, SiO43 and SiO42 displayed extended components exclusively in the blueshifted region relative to the stellar velocity, while SiO86 and SiO129 showed them only in the redshifted region. This discrepancy between the $J = 2-1$ and 3−2 masers and those in the $J=1-0$ state may be attributed to the differing spatial distributions of the preferred physical conditions among these masers. Another possible explanation could be the line overlap between the rovibrational transitions $v=2-1$, $J=1-0$ of $^{28}$SiO and $v_2=0-1$, $J_{k_a,k_c} = 12_{7,5} - 11_{6,6}$ of para-$H_2O$, as discussed in H. Olofsson et al. (1981) and R. Soria-Ruiz et al. (2004). However, our observations are insufficient to provide a definitive answer on the origin of these maser formations, as they were only detected in one light cycle during the monitoring period. Additionally, the observations were conducted using single-dish radio telescopes, which only offer information on the line-of-sight direction, not on the spatial distribution. Hence, as previously stated, further research on the spatial distribution of these masers is necessary.

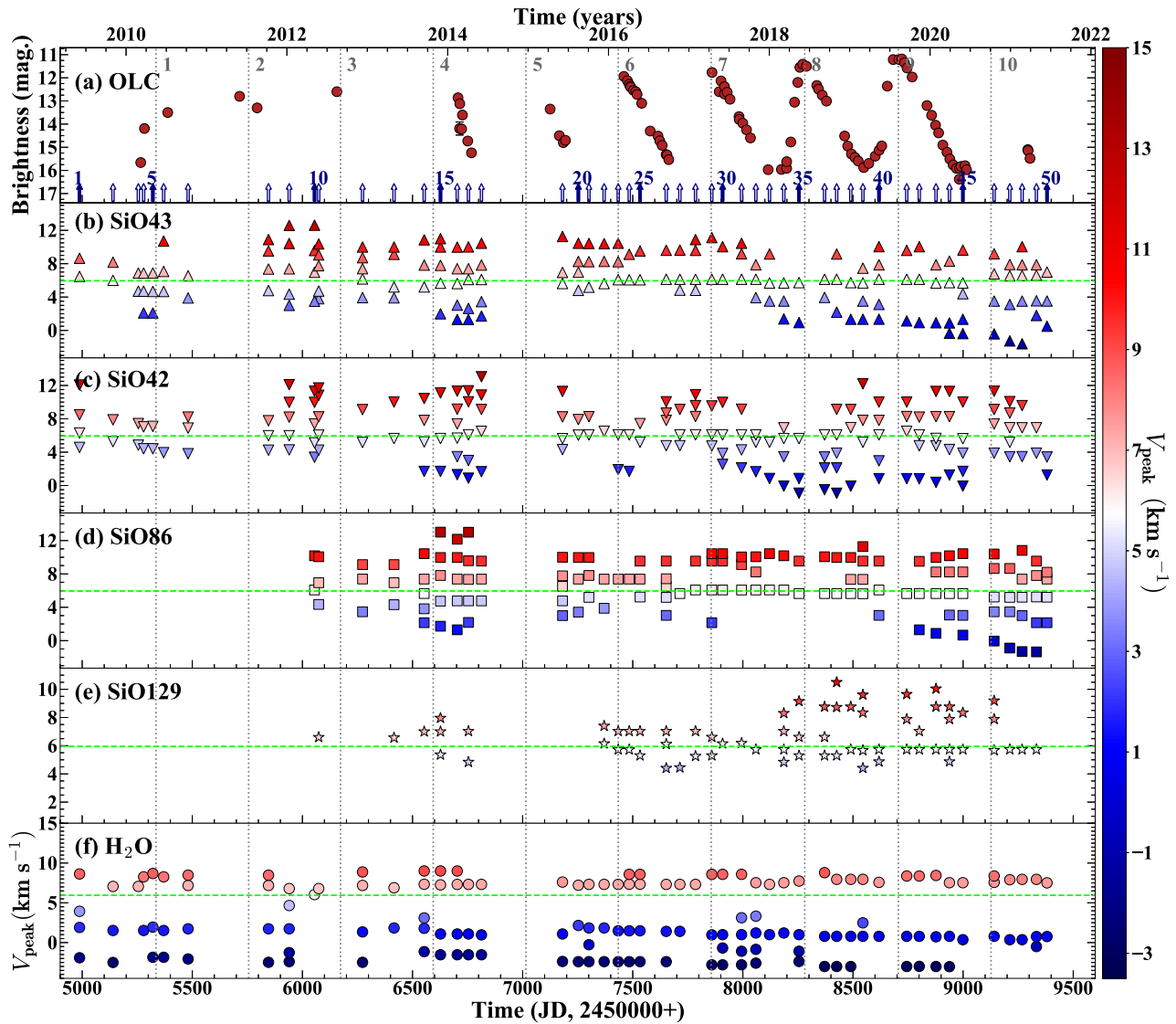

**Figure 7.** Peak velocity variations for all detected maser lines (panels (b)–(f)) relative to the OLC (panel (a)) over the period of 2009–2021. The vertical dotted lines represent the optical maximum phases. Green dashed lines indicate the $V_*$ for WX Ser. The color bar on the right side represents the range of peak velocities of masers.

It may be possible to use the time-averaged value of the measured $V_{mean}$ for the monitored SiO masers to deduce the $V_*$. Information on the stellar radial velocity/systemic velocity of AGB stars, mainly Mira variables and OH/IR stars, can be obtained by examining the velocity of thermal SiO or CO emissions or the central velocity of the double-peaked OH 1612 MHz line. Our analysis indicates that the time-averaged $V_{mean}$ of SiO masers could potentially serve as a marker for the $V_*$. While the $V_{peak}$ of SiO masers can be either red- or blueshifted in relation to the $V_*$, the $V_{mean}$ effectively tracks the $V_*$. This is because the $V_{mean}$ represents the emission of weighted velocity and can be viewed as the center of mass for the entire maser emission. The currently known $V_*$ of WX Ser is ~5.95 km s$^{-1}$, and the $V_{mean}$ averaged over time for SiO43, SiO42, SiO86, and SiO129 masers are 0.7, 0.8, 0.6, and 0.4 km s$^{-1}$ with respect to the $V_*$, respectively. Their velocity ranges are calculated to be 0.01–2.75 km s$^{-1}$, 0.01–3.20 km s$^{-1}$, 0.02–2.36 km s$^{-1}$, and 0.02–1.12 km s$^{-1}$ with respect to the $V_*$, respectively. Thus, the $V_{mean}$ of SiO129 is closest to the $V_*$ compared to the other SiO masers. From these results, we suggest that in future studies of targets with an unknown $V_*$, observations of SiO masers, if detected, will assist in tracing $V_*$.

As presented in Table 4 and Figures 1 and 7(f), the $H_2O$ maser of WX Ser exhibits a line profile primarily composed of four blueshifted components and two redshifted components, varying based on the observation epochs. Among these peaks, the B4, B2, R1, and R2 components were consistently detected throughout the monitoring period. As shown in Figure 2, the temporal variation of the velocity distribution of the $H_2O$ maser exhibits a markedly different pattern compared to that of the SiO masers during the same time frame. Notably, all observed SiO masers exhibit a kinematic distribution within a range that closely aligns with the $V_*$, whereas the velocity distribution of the $H_2O$ masers consistently expands relative to the $V_*$ throughout the observation period. In particular, it is evident that the velocities of the B4 and B2 components are escalating at a more rapid pace compared to the other components.

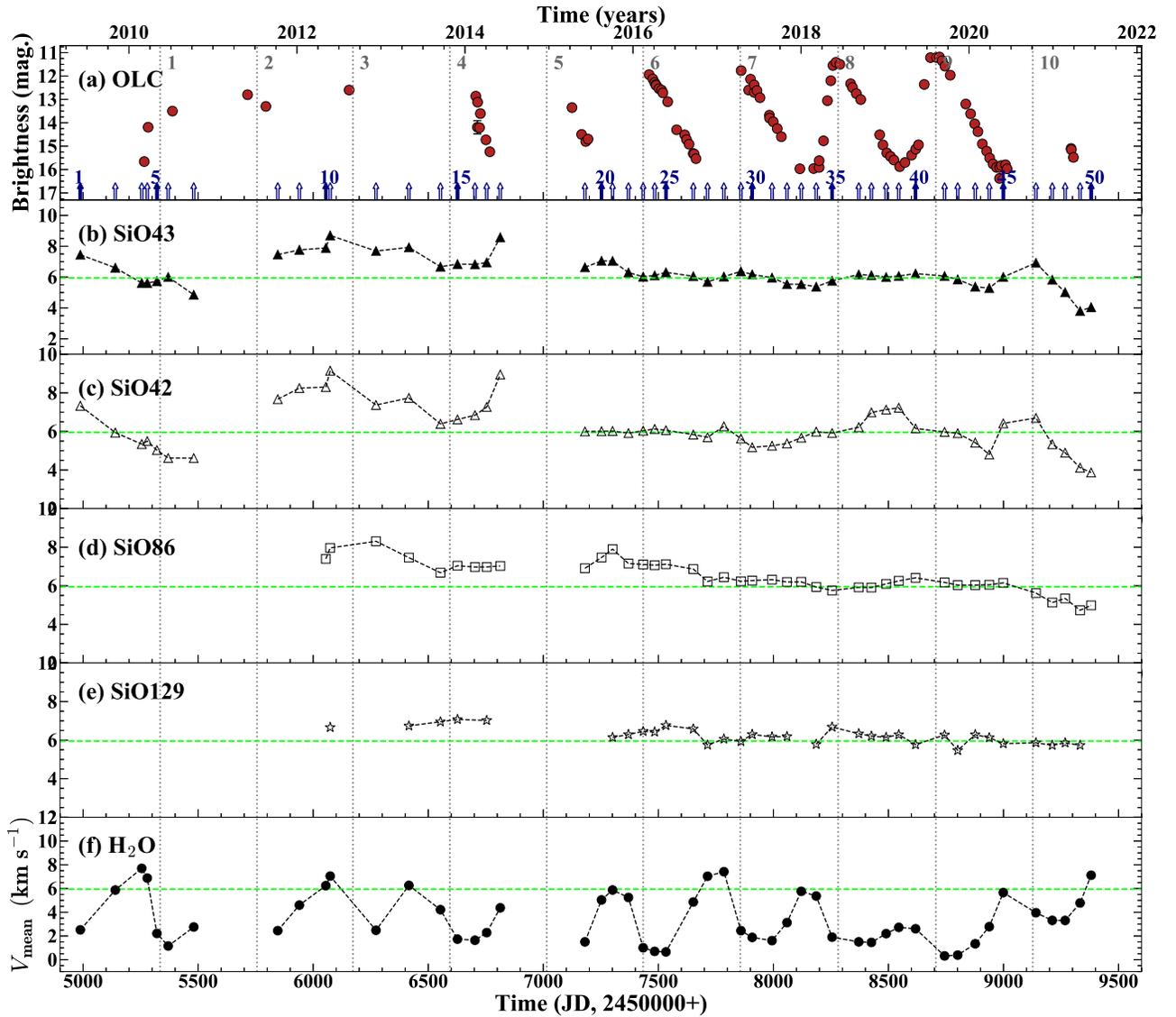

**Figure 8.** Mean velocity variation of all detected maser lines relative to the $V_*$ (panels (b)–(f)) and the OLC (panel (a)) over the period of 2009–2021. The vertical dotted lines indicate the optical maximum phases. Green dashed lines indicate the $V_*$ for WX Ser.

This trend is reflected in the time-varying mean velocity distribution of the $H_2O$ masers shown in Figure 8(f), which is relatively blueshifted. It is also noteworthy that the blueshifted concentration is pronounced at the optical maximum. These phenomena provide direct evidence that the material within the CSE undergoes acceleration and expansion due to the stellar wind generated by pulsation-driven shock waves. This supports the hypothesis that the ballistic motion of $H_2O$ masers is indicative of the considerable mass-loss process occurring in TP-AGB stars prior to their evolution into planetary nebulae. Furthermore, the asymmetry in the velocity distributions of the blue- and redshifted components, albeit for the line-of-sight velocity component, suggests that the spatial distribution of the $H_2O$ maser in this star is asymmetric as well.

In the meantime, the B2 component of most $H_2O$ masers is highly sensitive and strongly amplified at the optical maximum, as shown in Figures 2(e) and 4(f). Hence, the $V_{\rm mean}$ of the $H_2O$ maser is expected to be predominantly affected by the significant increase in flux density due to the density variation induced by stellar pulsation, particularly near the B2 peak. Figure 7(f) shows that the $V_{\rm peak}$ locations along B components are not always detected (e.g., epochs 3, 10, 11, and 28), and the peak intensities show significant variability. In contrast, the R-peaks are consistently identified even during the optical minimum phases. These results can be explained by a phenomenon in which "short-lived" clumps in the blueshifted velocity area repeatedly manifest in bursts, driven by stellar pulsation, whereas the "long-lived" clumps in the redshifted

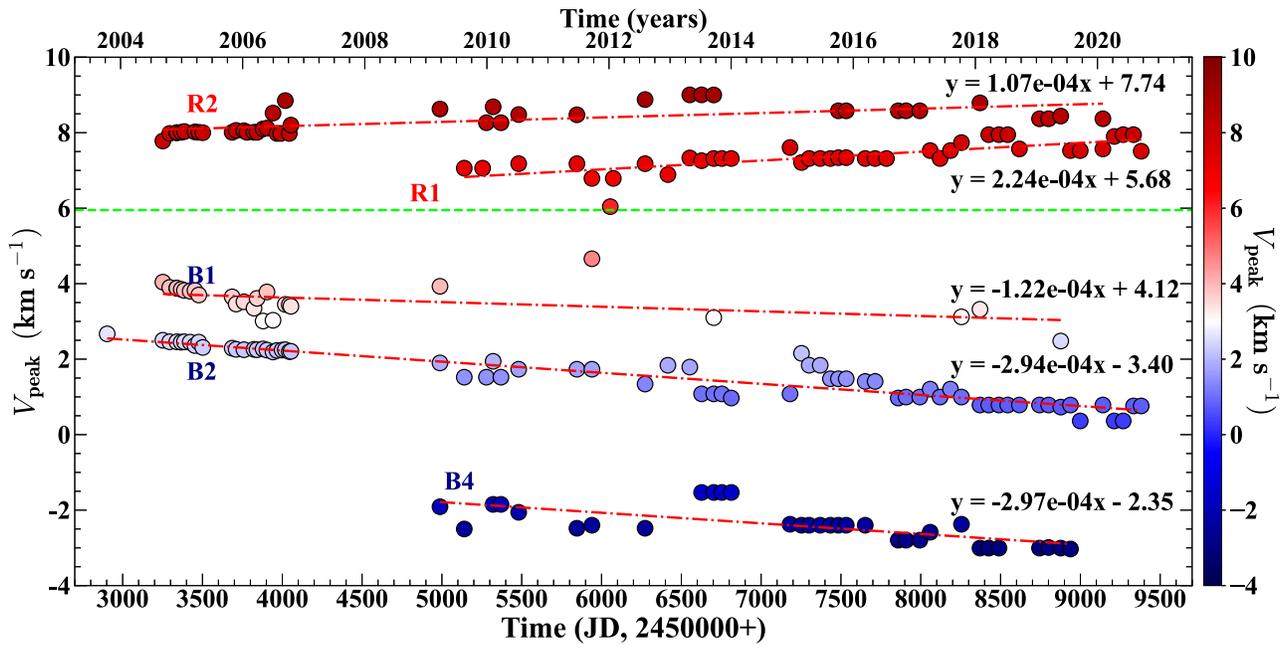

**Figure 9.** Time series of the H$_2$O maser peak velocities from 2003 to 2021. The observation data from 2003 to 2006 were extracted from the work of M. Shintani et al. (2008), while the data from 2009 to 2021 correspond to our own monitoring observations. The dashed green line represents the $V_*$, while the dashed–dotted red lines indicate inclinations of each component determined from the linear regression fitting method. A color scale is used to represent the magnitude of peak velocity. In the inserted equation, the unit of $x$ is days, while $y$ represents $V_{\rm peak}$.

velocity area are stable owing to increased density (J. Brand et al. 2020).

Additional observational data from M. Shintani et al. (2008) covering the period from 2003 to 2006 were taken into consideration in order to further investigate the long-term trend of the H$_2$O maser in WX Ser. We combined our observed data with data from M. Shintani et al. (2008) and performed a single linear regression fitting method for each component to identify long-term acceleration trends. The estimated functions are included in Figure 9 for reference. All components of the H$_2$O maser exhibit a gradual increase in velocity with respect to the $V_*$ over time, as shown in Figure 9. This provides direct evidence of the H$_2$O maser clumps' radial acceleration. In particular, the velocity gradients for the B1, B2, and B4 components are larger than those for the R1 and R2 components. The accelerations of the blueshifted components (B1, B2, and B4), as determined from the fitting results, are approximately −0.04, −0.11, and −0.11 km s$^{-1}$ yr$^{-1}$, respectively, while the accelerations of the redshifted components (R1 and R2) are ∼0.09 and ∼0.03 km s$^{-1}$ yr$^{-1}$, respectively.

This suggests that over the past 18 yr an asymmetric H$_2$O outflow has been detected in WX Ser. Several mechanisms can be proposed to account for this asymmetric H$_2$O outflow. One possibility is the existence of a binary companion, which can induce tidal interactions and trigger the outflow (O. De Marco et al. 2022). However, no companion object has been reported for WX Ser. Alternatively, a combination of stellar winds, magnetic fields, and the shock-driven multiphase instability may be responsible for launching and shaping the outflow. Another potential explanation may be the occurrence of intermittent mass ejection events during the TP-AGB stage (J. Kim et al. 2019). These events are characterized by the expulsion of nonuniform material from the stellar atmosphere, creating an asymmetric outflow. Our observational results support the notion that the initial bipolar or multipolar outflows of protoplanetary nebulae can begin to develop during the AGB phase, as previously proposed by A. A. Zijlstra et al. (2001) and B. Balick & A. Frank (2002). Recently, S. Xu et al. (2022) also demonstrated that the Mira variable BX Cam exhibits asymmetric outflows based on 3 yr of VLBI monitoring of its H$_2$O maser.

### 4.3.2. Full Width at Zero Intensity of SiO and H$_2$O Masers

The full width at zero intensity (FWZI) of masers provides useful information about the velocity distribution and kinematics of the expanding CSEs of AGB stars. By examining the FWZI, we can estimate the velocity at which the gas is moving away from the star, as well as identify any asymmetries or irregularities in the outflow, providing insights into the mass-loss mechanisms and evolutionary stages of these stars. In this context, we measured the FWZIs of the masers detected from WX Ser.

Table 7 presents the maximum blue- and redshifted velocities that delineate the edge of the emission line for each maser (from the first channel to the last channel where the emission is detected in the line profile). The FWZI is determined as the difference between the redshifted edge velocity ($V_{R,\rm edge}$) and blueshifted edge velocity ($V_{B,\rm edge}$): FWZI $= V_{R,\rm edge} - V_{B,\rm edge}$. The average values of FWZIs for the SiO43, SiO42, SiO86, SiO129, and H$_2$O masers were estimated to be 12.4, 12.3, 11.6, 7.8, and 16.2 km s$^{-1}$, respectively. Figure 10 shows the time-varying FWZIs of the five maser lines detected during our monitoring period. Linear fitting was performed on the FWZI values of each maser line, and the results are also shown in Figure 10.

For the SiO maser transition lines, there is a trend toward larger average FWZI values in the order of SiO129, SiO86, SiO42, and SiO43. It should be noted, however, that deviations among SiO43, SiO42, and SiO86 do not exceed 1 km s$^{-1}$. Taking into account the measurement error, the average FWZI values of these three maser transition lines are nearly identical, and the difference is negligible. However, it is evident that the

Table 7
Velocity Extents of SiO and H₂O Maser Lines (Kilometers per Second)

| Epoch | Date (yymmdd) | Phase ($\phi$) | SiO43 $V_{B,\text{edge}}$ | SiO43 $V_{R,\text{edge}}$ | SiO42 $V_{B,\text{edge}}$ | SiO42 $V_{R,\text{edge}}$ | SiO86 $V_{B,\text{edge}}$ | SiO86 $V_{R,\text{edge}}$ | SiO129 $V_{B,\text{edge}}$ | SiO129 $V_{R,\text{edge}}$ | H₂O $V_{B,\text{edge}}$ | H₂O $V_{R,\text{edge}}$ |
|---|---|---|---|---|---|---|---|---|---|---|---|---|
| 1 | 090605 | 0.18 | 2.2 | 14.3 | −0.2 | 12.6 | ⋯ | ⋯ | ⋯ | ⋯ | −3.2 | 10.8 |
| 2 | 091104 | 0.54 | 1.9 | 11.0 | 3.5 | 11.0 | ⋯ | ⋯ | ⋯ | ⋯ | −2.7 | 10.6 |
| 3 | 100227 | 0.81 | 2.1 | 11.1 | 2.0 | 10.3 | ⋯ | ⋯ | ⋯ | ⋯ | −2.4 | 11.0 |
| 4 | 100323 | 0.87 | 1.2 | 11.2 | 2.2 | 8.6 | ⋯ | ⋯ | ⋯ | ⋯ | −2.5 | 10.0 |
| 5 | 100504 | 0.97 | 1.7 | 12.2 | 2.7 | 9.6 | ⋯ | ⋯ | ⋯ | ⋯ | −3.7 | 12.2 |
| 6 | 100622 | 1.08 | −0.1 | 13.8 | 1.7 | 8.4 | ⋯ | ⋯ | ⋯ | ⋯ | −5.5 | 11.0 |
| 7 | 101012 | 1.35 | 2.2 | 10.2 | −1.0 | 10.0 | ⋯ | ⋯ | ⋯ | ⋯ | −6.5 | 11.7 |
| 8 | 111012 | 2.22 | 1.3 | 14.0 | −0.2 | 14.8 | ⋯ | ⋯ | ⋯ | ⋯ | −4.8 | 10.4 |
| 9 | 120103 | 2.44 | 0.7 | 14.1 | 2.1 | 13.9 | ⋯ | ⋯ | ⋯ | ⋯ | −6.5 | 12.9 |
| 10 | 120507 | 2.72 | 2.2 | 17.4 | 2.5 | 14.0 | 3.0 | 11.3 | ⋯ | ⋯ | 2.4 | 9.8 |
| 11 | 120526 | 2.76 | 2.1 | 13.1 | 1.6 | 14.4 | 3.4 | 12.5 | 4.3 | 8.1 | 1.9 | 15.0 |
| 12 | 121211 | 3.24 | 2.6 | 13.1 | 2.5 | 13.5 | 2.1 | 13.9 | ⋯ | ⋯ | −5.7 | 14.6 |
| 13 | 130503 | 3.58 | 2.2 | 15.2 | 2.1 | 11.8 | 2.2 | 11.4 | 5.1 | 7.9 | −3.4 | 11.3 |
| 14 | 130917 | 3.90 | 1.7 | 14.0 | 0.4 | 13.6 | 0.8 | 12.5 | 4.7 | 10.8 | −3.9 | 10.9 |
| 15 | 131130 | 4.08 | 1.3 | 15.2 | −0.1 | 12.6 | 0.9 | 15.6 | 1.7 | 10.3 | −5.1 | 12.2 |
| 16 | 140214 | 4.26 | −0.5 | 12.6 | −1.4 | 14.0 | −0.6 | 14.8 | 4.7 | 8.9 | −5.1 | 10.9 |
| 17 | 140406 | 4.38 | −3.1 | 14.8 | −1.4 | 14.9 | 0.4 | 13.4 | 3.3 | 9.8 | −4.7 | 12.2 |
| 18 | 140604 | 4.52 | 1.9 | 13.6 | 0.7 | 15.2 | 3.1 | 12.6 | ⋯ | ⋯ | −5.1 | 14.4 |
| 19 | 150607 | 5.40 | 1.7 | 13.0 | 3.0 | 11.9 | 1.3 | 12.6 | ⋯ | ⋯ | −6.0 | 13.4 |
| 20 | 150818 | 5.57 | 1.8 | 13.1 | 3.4 | 13.0 | 3.1 | 12.5 | ⋯ | ⋯ | −5.7 | 13.7 |
| 21 | 151005 | 5.68 | 2.2 | 13.1 | 2.5 | 12.1 | 3.4 | 11.4 | ⋯ | ⋯ | −3.0 | 9.2 |
| 22 | 151213 | 5.85 | 1.7 | 12.3 | 2.1 | 14.8 | 0.9 | 12.6 | 3.3 | 11.1 | −3.0 | 10.1 |
| 23 | 160215 | 6.00 | 0.9 | 12.6 | −0.4 | 11.0 | 2.5 | 11.3 | 4.2 | 9.9 | −5.1 | 10.5 |
| 24 | 160405 | 6.12 | 0.4 | 13.9 | 1.2 | 11.3 | 1.3 | 13.0 | 2.5 | 10.3 | −5.9 | 11.5 |
| 25 | 160524 | 6.23 | 0.3 | 15.7 | 0.8 | 14.9 | 0.0 | 11.3 | 2.5 | 9.9 | −5.9 | 13.1 |
| 26 | 160920 | 6.52 | 2.2 | 11.0 | 2.6 | 11.0 | 2.6 | 14.3 | 2.1 | 10.7 | −6.4 | 10.1 |
| 27 | 161121 | 6.66 | 2.2 | 11.0 | 1.7 | 14.0 | 2.6 | 10.9 | 2.1 | 9.5 | −3.4 | 9.6 |
| 28 | 170131 | 6.83 | 1.8 | 12.3 | 1.6 | 14.0 | 2.2 | 10.9 | 1.2 | 10.7 | −0.5 | 13.4 |
| 29 | 170415 | 7.01 | 0.9 | 13.6 | −0.1 | 12.2 | 2.2 | 11.0 | 2.1 | 10.3 | −6.8 | 13.4 |
| 30 | 170603 | 7.12 | 1.7 | 12.2 | −1.8 | 14.3 | 2.5 | 11.7 | 0.3 | 11.5 | −7.6 | 10.9 |
| 31 | 170828 | 7.32 | 1.3 | 14.0 | −0.4 | 12.3 | 2.6 | 10.9 | 0.3 | 9.0 | −5.6 | 12.2 |
| 32 | 171101 | 7.48 | −0.3 | 12.2 | −3.2 | 10.5 | 1.7 | 11.3 | 4.2 | 11.6 | −4.9 | 10.7 |
| 33 | 180102 | 7.62 | −1.7 | 11.5 | −1.0 | 10.1 | 2.6 | 11.3 | ⋯ | ⋯ | −1.3 | 12.2 |
| 34 | 180308 | 7.78 | −0.8 | 10.5 | −1.4 | 10.0 | −1.3 | 11.3 | 3.7 | 11.1 | −2.4 | 12.0 |
| 35 | 180516 | 7.94 | −0.9 | 10.5 | −2.3 | 11.0 | 0.4 | 13.4 | 2.5 | 14.6 | −4.0 | 10.9 |
| 36 | 180909 | 8.21 | −1.3 | 11.8 | −2.2 | 11.8 | −1.0 | 11.7 | 3.3 | 13.4 | −5.3 | 14.9 |
| 37 | 181103 | 8.34 | −3.0 | 13.9 | −3.6 | 11.4 | −1.0 | 12.2 | 1.6 | 12.9 | −4.5 | 10.7 |
| 38 | 190106 | 8.49 | −0.9 | 10.5 | −2.2 | 13.5 | −0.8 | 12.6 | 2.9 | 11.6 | −4.9 | 13.2 |
| 39 | 190302 | 8.62 | −1.7 | 11.9 | −4.0 | 12.7 | 0.8 | 13.9 | 4.2 | 10.7 | −3.5 | 10.3 |
| 40 | 190514 | 8.79 | −0.8 | 11.9 | −2.3 | 14.3 | 0.0 | 13.1 | 3.7 | 10.7 | −3.6 | 10.8 |
| 41 | 190917 | 9.09 | −0.3 | 12.7 | −1.4 | 13.1 | −1.0 | 11.7 | 1.6 | 12.0 | −6.2 | 11.1 |
| 42 | 191113 | 9.23 | −1.8 | 14.0 | −1.9 | 13.6 | 0.5 | 13.8 | 4.2 | 9.9 | −8.2 | 10.7 |
| 43 | 200128 | 9.41 | −0.8 | 12.7 | −1.0 | 13.0 | −2.6 | 11.7 | 3.7 | 12.0 | −8.0 | 10.8 |
| 44 | 200329 | 9.55 | −0.8 | 9.7 | −0.5 | 14.9 | −0.1 | 13.8 | 4.2 | 12.0 | −8.7 | 11.6 |
| 45 | 200529 | 9.70 | −1.2 | 13.5 | −1.0 | 13.6 | −0.9 | 11.7 | 3.3 | 9.5 | −1.1 | 14.5 |
| 46 | 201018 | 10.03 | −1.3 | 12.2 | 0.8 | 12.2 | −2.2 | 12.1 | 4.2 | 13.4 | −2.4 | 16.6 |
| 47 | 201228 | 10.20 | −2.6 | 11.4 | −2.3 | 12.2 | −3.1 | 11.4 | 2.1 | 10.7 | −5.3 | 11.1 |
| 48 | 210222 | 10.33 | −2.1 | 11.0 | −2.2 | 10.0 | −2.2 | 13.6 | 3.7 | 10.7 | −5.3 | 10.3 |
| 49 | 210428 | 10.49 | −0.4 | 9.2 | 0.3 | 10.8 | −2.3 | 11.4 | 3.3 | 11.6 | −5.3 | 12.9 |
| 50 | 210615 | 10.60 | 0.1 | 10.1 | −1.9 | 9.2 | 0.3 | 11.4 | ⋯ | ⋯ | −2.8 | 11.6 |

(This table is available in machine-readable form in the online article.)

average FWZI values for SiO129 are notably lower when compared to the FWZI values measured for the other three maser transitions. The likely reason for this discrepancy is the relatively lower sensitivity to SiO129 observations compared to SiO43, SiO42, and SiO86. The average rms noise levels for SiO43, SiO42, and SiO86 are 0.21, 0.21, and 0.27 Jy, respectively. In contrast, the average rms noise for SiO129 stands at 0.37 Jy, approximately 1.6 times higher than the other three masers as detailed in Table 3.

The FWZI values of SiO43 and SiO42 have been statistically analyzed in previous studies for a variety of Mira variable stars. Specifically, J. Kim et al. (2014) conducted measurements on approximately 200 Mira variable stars, expressing FWZI as FWZP in their paper. The calculated sample means for the FWZIs of SiO43 and SiO42 were found to be 13.7 and 13.0 km s$^{-1}$, respectively. However, a few outliers were identified within their sample, prompting us to recalculate the 10%-trimmed mean by appropriately excluding these data points. The recalculated mean values were determined to be 13.3 and 12.6 km s$^{-1}$ for SiO43 and SiO42, respectively, with corresponding ranges of 7–22 km s$^{-1}$ and 7–21 km s$^{-1}$. For WX Ser, the average FWZIs for SiO43 and SiO42 were calculated to be 12.4 and 12.3 km s$^{-1}$, respectively, with ranges of 8–18 km s$^{-1}$ and 6–17 km s$^{-1}$, respectively. These values fall within the above ranges typically identified as Mira variable stars.

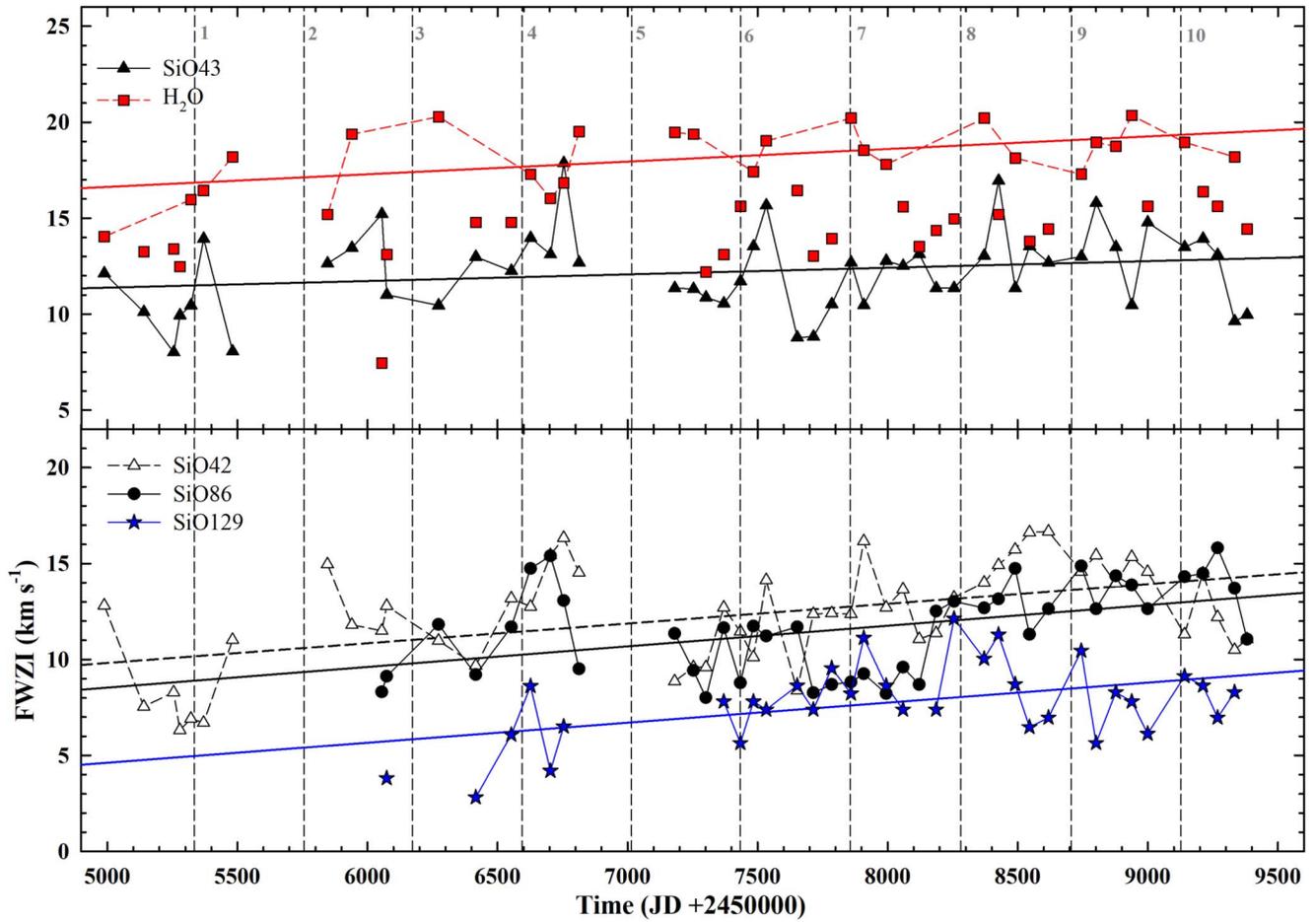

**Figure 10.** Temporal variations of the FWZIs of SiO and H$_2$O maser lines. The vertical dashed lines indicate the optical maximum phases. The symbols used are presented in the legend inside the plot. Each line represents the result of linear fitting to the FWZI of each maser. The red solid line represents the linear fitting line for the FWZI of the H$_2$O maser, derived from the values of the B4 and R2 components (see text).

In the case of the H$_2$O maser, we focused on identifying a clear increase in velocity by specifically measuring the FWZI values for the B4 and R2 components. These components exhibit the most significant velocity differences compared to the $V_*$, as shown in Figures 2(f) and 9. The FWZIs of these values show a general increasing trend over time, ranging from approximately 14.0 to 20.3 km s$^{-1}$, with an average FWZI value of 18.2 km s$^{-1}$ for these components alone. These results are depicted in Figure 10, along with the results of linear fitting. The FWZI of the H$_2$O maser from these two components is significantly larger than those of the SiO masers. This is likely related to the different placements of these masers within the stellar atmosphere and outer shell. SiO masers are mainly detected in a region located in close proximity to the stellar surface, where the gaseous material retains a significant level of density and initiates its outward expulsion. Since SiO masers are positioned closer to the star, they have to account for both the material's turbulence and the star's gravity. As a result, they will inherently have slower expansion velocities than H$_2$O masers located at greater distances. On the other hand, H$_2$O masers are typically found in the outer regions of the stellar envelope. In these areas, the material that has been ejected from the star is propelled outward by a combination of radiation pressure and other forces, ultimately reaching its terminal velocity. In particular, this difference in WX Ser may be attributable to the peculiar characteristics of the H$_2$O maser in this star, which exhibits an asymmetric outflow dominated by blueshifted velocities. Furthermore, the outward expansion of the H$_2$O maser accelerates over time. This result confirms the hypothesis that the massive expulsion of stellar material during the TP-AGB phase, before the transition to the planetary nebula phase, can be traced through the ballistic movement of H$_2$O masers.

If we consider half of the FWZI of the H$_2$O maser's two components (B4 and R2) as an indicator of the expansion velocity of the CSE, an estimated value of 9.1 km s$^{-1}$ is obtained. On the other hand, the terminal velocity of WX Ser, as determined from OH maser observations, is 7.6 km s$^{-1}$ (J. Herman & H. J. Habing 1985). This discrepancy arises because the H$_2$O maser region exhibits an ongoing outward acceleration due to the significant mass ejection occurring within the CSE, reflecting the initial ballistic movement. In addition, the H$_2$O maser detected in WX Ser exhibits a highly asymmetric expansion pattern with a notable bias toward the blueshifted velocity. Considering the blueshifted velocity of the H$_2$O maser relative to the stellar velocity, i.e., $|V_{B,\mathrm{edge}} - V_*|$, the maximum recorded value is 14.7 km s$^{-1}$, which is comparable to that of OH/IR stars.

The presence of a high-velocity asymmetric outflow structure in a low-mass AGB star, such as WX Ser, indicates that the expanding H$_2$O maser cloud is highly active during the Mira variable phase. As the evolutionary process advances and the ejection of mass from the central star ceases, the resulting accelerated expansion will exhibit characteristics of isokinetic motion. This phenomenon will lead to the initial development of a bipolar planetary nebula.

## 5. Summary and Conclusions

We performed a long-term simultaneous monitoring observation of SiO and H$_2$O masers pertaining to the Mira variable star WX Ser using 21 m single-dish telescopes of the KVN. The observations spanned from 2009 June to 2021 June and focused on five maser lines: SiO $v = 1, 2, J = 1-0$ (43.1 and 42.8 GHz); SiO $v = 1, J = 2-1, J = 3-2$ (86.2 and 129.3 GHz); and H$_2$O $6_{1,6}-5_{2,3}$ (22.2 GHz).

The line profiles of the SiO and H$_2$O masers in this star present a complex and intriguing picture. The SiO maser line profiles exhibit a concentration near the $V_*$, yet they also reveal multiple components, ranging from five to six depending on the optical phase. In contrast, the line profiles of the H$_2$O maser are characterized by blueshifted velocity-biased dominant emission lines, showcasing up to six components in total. This nuanced behavior underscores the intricate dynamics at play within the maser emissions of this star. In particular, the presence of the biased H$_2$O emission leads to the manifestation of various asymmetries in the CSE. In the meantime, we detected a sudden and intense flaring event of the H$_2$O maser emission with $\sim$120 Jy in 2019 September ($\phi = 9.09$ out of 10 cycles).

The intensity variation patterns of SiO and H$_2$O masers are strongly correlated with the OLC of the central star. However, phase lags exist between the optical maxima of the central star and intensity maxima of the SiO and H$_2$O masers. The phase lag observed between the H$_2$O maser and the OLC often exhibits a more pronounced difference than the phase lag between the SiO maser and the OLC. However, this difference is somewhat dependent on changes in the optical period. Furthermore, there is a noticeable, albeit slight, phase lag between the SiO and H$_2$O masers, mainly attributed to the distinct formation regions and the different driving mechanisms that pump the masers.

Moreover, we examined the temporal variations in the peak and mean velocities of five SiO and H$_2$O masers. The velocity variation trends of the SiO masers are similar to those of other Mira variable stars. However, the variation pattern of the H$_2$O maser is significantly different from that of the SiO masers. The H$_2$O maser exhibits a consistent radial acceleration over time, and the overall pattern of the mean velocity of the H$_2$O maser indicates a predominant bias toward the blueshifted velocity.

Time-varying intensity and velocity morphologies may be a result of asymmetric outflow caused by nonuniform material ejection from the stellar atmosphere. After integrating all of the previous data from 2003 onward, we found that the expansion velocity of the H$_2$O maser is increasing. After being identified in order, five H$_2$O maser features (B1, B2, B4, R1, R2) were found, with estimated accelerations of 0.04, 0.11, and 0.11 km s$^{-1}$ yr$^{-1}$ in the blueshifted direction and 0.09 and 0.03 km s$^{-1}$ yr$^{-1}$ in the redshifted direction, respectively. The B1, B2, and B4 components accelerate at a slightly higher rate than the R1 and R2 components in relation to $V_*$. In other words, the blueshifted asymmetric outflow of the H$_2$O maser has been developing in WX Ser for at least 18 yr. These velocity characteristics suggest the emergence of an initial asymmetric outflow structure during the TP-AGB phase, specifically in the Mira variable star stage.

The detection of an asymmetric H$_2$O outflow in the Mira variable star WX Ser has important implications for our understanding of stellar evolution and the formation of bipolar and/or multipolar planetary nebulae. By studying the dynamics and properties of these initial outflows, we can gain valuable insights into the mechanisms that transform the spherical CSEs of low- and intermediate-mass evolved stars into nonspherical shapes. To further unravel the mysteries of WX Ser, further observations are necessary. For example, high-resolution imaging observations can provide more detailed information regarding the structure and kinematics of the outflow.

### Acknowledgments

We thank the anonymous referee of this paper for very helpful comments. The KVN observations were made possible by the high-speed network connections among the KVN sites provided by the Korea Research Environment Open NETwork (KREONET), which is managed and operated by the Korea Institute of Science and Technology Information (KISTI). Moreover, we appreciate the variable star observations provided by the AAVSO International Database, which have been contributed by observers worldwide and were used in this research. The work described in this paper was supported by a grant from KASI (project No. 2023-1-840-00) and grants from the National Research Foundation of Korea (NRF) (J.-H.L, H.K: No. NRF-2021R1A2C1008928; S.-H. C.: No. NRF-2022R1A2C1091317; D.-H.Y: No. NRF-2019 R1A6A3A01091901; K.-W.S: No. NRF-2022R1I1A3055131), funded by the Korean government through the Ministry of Science and ICT (MSIT).

*Facility:* KVN.

*Software:* GILDAS (J. Pety 2018).

### ORCID iDs

Jang-Ho Lim https://orcid.org/0009-0004-6773-5754
Jaeheon Kim https://orcid.org/0000-0001-9825-7864
Se-Hyung Cho https://orcid.org/0000-0002-2012-5412
Hyosun Kim https://orcid.org/0000-0001-9639-0354
Dong-Hwan Yoon https://orcid.org/0000-0001-7120-8851
Seong-Min Son https://orcid.org/0009-0007-1400-7413
Kyung-Won Suh https://orcid.org/0000-0001-9104-9763